\documentclass[11pt]{article}
\usepackage{amsfonts, amsmath, amssymb, amsthm}
\usepackage[english]{babel}
\usepackage{booktabs}
\usepackage{bm}
\usepackage{eucal}
\usepackage{enumerate}
\usepackage{float}
\usepackage{dsfont} 
\usepackage{epsfig}
\usepackage {fontenc}
\usepackage[hmargin=1.6cm,vmargin=3cm]{geometry}
\usepackage{lscape}
\usepackage{mathrsfs}
\usepackage{natbib}
\usepackage{rotating}
\usepackage{verbatim}
\usepackage{xcolor}
\usepackage{dirtytalk}
\usepackage{multirow}
\usepackage{mdframed}
\usepackage[normalem]{ulem}
\usepackage{subfigure}
\usepackage{xr}
\usepackage{mleftright}
\usepackage{hyperref}

\date{}

\renewcommand{\thefootnote}{$\dagger$}

\begin{document}
\title{Density regression via Dirichlet process mixtures of normal structured additive regression models}
\author{\textsc{Mar\'ia Xos\'e Rodr\'iguez-\'Alvarez}, \textsc{Vanda~In\'acio}, and \textsc{Nadja Klein}}
\date{}
\maketitle 
\begin{abstract}
There is an increased interest in studying how the distribution of a continuous response variable changes with a set of covariates. Within a Bayesian nonparametric framework, dependent Dirichlet process mixture models provide a highly flexible approach for conducting inference about the conditional density function. However, several formulations of this class make either rather restrictive modelling assumptions or involve intricate  algorithms for posterior inference, thus preventing their widespread use. In response to these challenges, we present a flexible, versatile, and computationally tractable model for density regression based on a single-weights dependent Dirichlet process mixture of normal distributions model for univariate continuous responses. We assume an additive structure for the mean of each mixture component and incorporate the effects of continuous covariates through smooth nonlinear functions. The key components of our modelling approach are penalised B-splines and their bivariate tensor product extension. Our proposed method also seamlessly accommodates parametric effects of categorical covariates, linear effects of continuous covariates, interactions between categorical and/or continuous covariates, varying coefficient terms, and random effects, which is why we refer our model as a Dirichlet process mixture of normal structured additive regression models. A noteworthy feature of our method is its efficiency in posterior simulation through Gibbs sampling, as closed-form full conditional distributions for all model parameters are available. Results from a simulation study demonstrate  that our approach successfully recovers true conditional densities and other regression functionals in various challenging scenarios. Applications to a toxicology, disease diagnosis, and agricultural study are provided and further underpin the broad applicability of our modelling framework. An \texttt{R} package, \texttt{DDPstar}, implementing the proposed method is publicly available at \url{https://bitbucket.org/mxrodriguez/ddpstar}.
\end{abstract}

\let\thefootnote\relax\footnotetext{Mar\'ia Xos\'e Rodr\'iguez-\'Alvarez, Departamento de Estat\'istica e Investigaci\'on Operativa, Universidade de Vigo, Vigo, Spain (\textit{mxrodriguez@uvigo.gal}). Vanda In\'acio, School of Mathematics, University of Edinburgh, Scotland, UK (\textit{vanda.inacio@ed.ac.uk}). Nadja Klein, Chair of Uncertainty Quantification and Statistical Learning, Research Center
Trustworthy Data Science and Security (UA Ruhr) and Department of Statistics
(Technische Universitat Dortmund), Dortmund, Germany (\textit{nadja.klein@tu-dortmund.de}).}

\textsc{key words:} Bayesian nonparametrics; Bivariate smoothing; Conditional density; Gibbs sampling; Penalised B-splines

\section{\large{\textsf{Introduction}}}
In this article we are concerned about developing a flexible model for density regression that allows investigating how the distribution of a univariate continuous, real-valued, response variable $y$ changes as a function of covariates  $\boldsymbol{x}\in\mathcal{X}\subset \mathbb{R}^{p}$.

The literature on Bayesian flexible models for mean regression is vast and include, among others, (penalised) splines \citep{Gustafson2000, Lang2004}, Gaussian processes \citep[][Chapter 2]{Williams2006}, neural networks \citep[][Chapter 5]{Bishop2006}, and (Bayesian) additive regression trees \citep{Chipman2010}. These methods, although having the potential to approximate a wide range of regression functions, only allow for flexibility in the conditional mean of $y$, assuming that the scale and higher order moments are not affected by $\boldsymbol{x}$. Extensions to Bayesian location-scale regression models where both the mean and variance are covariate-dependent and flexibly modelled have been considered as well \citep[e.g.,][]{Rodriguez2014,Pratola2020}. 
An alternative to location(-scale) regression models is quantile regression, which directly models a conditional quantile of the response variable as a function of the covariates. Bayesian flexible approaches for this class of models include those developed, for instance, by 
\cite{Reich2010} and \cite{Waldmann2013}. When the goal is to extend beyond simply modelling specific functionals of the response distribution, such as the mean, variance, or a quantile, through covariates, Bayesian distributional regression models \citep{KleKneLanSoh2015} arise as a natural option. This class of models, also known as Bayesian additive models for location, scale, and shape, is able to provide a complete probabilistic characterisation of the response distribution by making a parametric assumption on the conditional density and by potentially relating each parameter of such a density to an additive predictor. However, choosing an appropriate parametric form for the conditional density may not be a trivial task in many applications. Furthermore, the underlying assumption that the same parametric distribution family applies over the whole covariate space may be overly restrictive.

Within the Bayesian nonparametric literature, several methods relying on generalisations of mixture models for marginal density estimation have been proposed that provide flexible inference for conditional densities. This class of models does not require a specific parametric assumption about the conditional response distribution; it `only' necessitates the specification of a family of distributions for the mixture components. A key advantage is that these methods can easily handle intricate distributional features, such as multimodality, skewness, and/or extreme variability, without the need to know of their existence in advance. Compared to regression models that focus on specific functionals, Bayesian nonparametric density regression yields coherent inference and allows to derive any quantity of interest, such as the conditional mean, variance, or quantiles. Furthermore, it has the advantage over many quantile regression approaches in that the monotonicity constraint is trivially satisfied, thus preventing quantiles from crossing.

In what regards marginal density estimation, Dirichlet process mixture models are an extremely useful tool as they combine the attractive features of mixture modelling with theoretical properties of nonparametric priors, such as full support and posterior consistency \citep[see, e.g.,][Chapter 7]{Ghosal2017}. These models are of the form
\begin{equation}\label{dpm}
p(y\mid G) \equiv p(y) = \int k(y\mid\boldsymbol{\theta})\text{d}G(\boldsymbol{\theta}),\qquad G\sim\text{DP}(\alpha,G_0(\boldsymbol{\theta})),
\end{equation}
where $k(\cdot\mid\boldsymbol{\theta})$ is the density for the mixture kernel family of distributions with parameter $\boldsymbol{\theta}$ and the mixing distribution $G$ follows a Dirichlet process (DP) prior \citep{Ferguson1973} with centring distribution $E(G) = G_0(\boldsymbol{\theta})$ and precision parameter $\alpha>0$. The stick-breaking representation of the DP \citep{Sethuraman1994}, under which $G$ can be written as an infinite weighted sum of point masses
\begin{equation}\label{sethu}
G(\cdot)=\sum_{l=1}^{\infty}\omega_l\delta_{\boldsymbol{\theta}_l}(\cdot),\quad \omega_l=\begin{cases} \eta_1, & \text{if } l=1, \\ \eta_l\prod_{m<l}(1-\eta_m), & \text{if } l > 1,
\end{cases}\quad \eta_l\overset{\text{iid}}\sim\text{Beta}(1,\alpha),\quad \boldsymbol{\theta}_l\overset{\text{iid}}\sim G_0,\quad l\geq 1,
\end{equation}
further allows to write the density in \eqref{dpm} as a countable mixture of kernel densities
\begin{equation*}
p(y)=\sum_{l=1}^{\infty}\omega_lk(y\mid \boldsymbol{\theta}_l),
\end{equation*}
where the collections $\{\omega_l\}_{l\geq 1}$ and $\{\boldsymbol{\theta}_l\}_{l\geq 1}$ are independent of each other. When the kernel is a normal density, i.e.,  $k(y\mid\boldsymbol{\theta})=\phi(y\mid\mu,\sigma^2)$, $\boldsymbol{\theta}=(\mu,\sigma^2)$, the resulting model has been shown to approximate any smooth density on the real line \citep{Lo1984}. DP mixture models are well-established nowadays and a variety of samplers for efficient posterior simulation are available \citep[e.g.,][]{Neal2000,Ishwaran2001,Kalli2011}. 

When covariate information is available, a model for the collection of conditional densities $\{p(y\mid\boldsymbol{x}): \boldsymbol{x}\in\mathcal{X}\subset \mathbb{R}^{p}\}$ can be obtained by allowing the mixing distribution $G$ in \eqref{dpm} to depend on $\boldsymbol{x}$, i.e., 
\begin{equation*}
p(y\mid G_{\boldsymbol{x}}) \equiv p(y\mid\boldsymbol{x})=\int k(y\mid\boldsymbol{\theta})\text{d}G_{\boldsymbol{x}}(\boldsymbol{\theta}).
\end{equation*}
One possibility for a prior on the collection of mixing distributions $\{G_{\boldsymbol{x}}: \boldsymbol{x}\in\mathcal{X}\}$ is the dependent Dirichlet process (DDP) proposed by \cite{Maceachern1999,Maceachern2000}, which builds upon the stick-breaking representation of the DP in \eqref{sethu} and which in its full generality makes both the weights and the atoms dependent on covariates, that is, $G_{\boldsymbol{x}}(\cdot) = \sum_{l=1}^{\infty}\omega_l(\boldsymbol{x})\delta_{\boldsymbol{\theta}_{l}(\boldsymbol{x})}(\cdot)$. Here $\{\eta_l(\boldsymbol{x})\}_{l\geq 1}$, the inputs of the weights $\omega_l(\boldsymbol{x})$ in $G_{\boldsymbol{x}}$, and $\{\boldsymbol{\theta}_l(\boldsymbol{x})\}_{l\geq 1}$, the atoms of the mixture model, are collections of stochastic processes defined on $\mathcal{X}$ and are independent of each other. The key idea behind the DDP construction is that at each $\boldsymbol{x} \in \mathcal{X}$, $G_{\boldsymbol{x}}$ is marginally a DP. An in-depth review of DDPs and related models is provided in \cite{Quintana2022} and \cite{Wade2023}. In brief, most proposals in the literature fall into either: (i) models with covariate-dependent weights and atoms that are either constant across the values of $\boldsymbol{x}$ or that rely on a linear formulation \citep[see, among others,][although we note that some of these correspond to a variation of MacEachern's original DDP where, marginally, $G_{\boldsymbol{x}}$ does not necessarily correspond to a DP]{Griffin2006, Dunson2007,Dunson2008,Chung2009,Antoniano2014,Rigon2021}, and (ii) models with covariate-dependent atoms but common (single) weights across the values of $\boldsymbol{x}$, i.e., $\omega_l(\boldsymbol{x}) = \omega_l$ \citep[see, for instance,][]{deiorio2004,Gelfand2005,delacruz2007,Rodriguez2008,deiorio2009,Jara2010,Gutierrez2011,Fronczyk2014,Xu2016,Xu2019,Xu2022}.


Formulations that rely on covariate-dependent weights are very flexible, but computations, as highlighted in \citet[][p.~37]{Quintana2022}, \citet[][p.~132]{Rigon2021}, and \citet[][p.~12]{Wade2023}, tend to be burdensome. An exception to the cumbersome computations issue is the logit stick-breaking prior of \cite{Rigon2021}, which is computationally tractable and posterior inference is available under several computational schemes. Additionally, the stick-breaking definition poses challenges in terms of the different choices that need to be made for functional shapes and hyperparameters when defining the weights' inputs $\eta_l(\boldsymbol{x})$. These challenges are amplified by the lack of interpretation of the quantities involved (\citealt[p.~478]{Antoniano2014}, \citealt[p.~23]{Wade2023}).

In contrast, the single-weights DDP model is very popular, mainly due to the ease of prior specification and its computational simplicity, as posterior simulation can be implemented using the same sampling algorithms available for  DP mixtures. However, although such class of models, under its most general formulation, has desirable theoretical properties \citep{Barrientos2012,Pati2013}, it may have limited flexibility in terms of the regression relationships it can capture (see more in Section \ref{sec:model}). Nevertheless, quoting \citet[p.~16]{Maceachern2000}: ``\emph{They (single-weights DDPs) thus provide a general framework that covers a vast territory}''.

In this article we develop a single-weights DDP mixture of normal distributions model that overcomes the aforementioned lack of flexibility but retains the computational tractability. Specifically, we define the atoms of each normal component as $\boldsymbol{\theta}_{l}(\boldsymbol{x})=(\mu_l(\boldsymbol{x}),\sigma_l^2)$ and assume an additive structure for $\mu_{l}(\boldsymbol{x})$, allowing the incorporation of different types of covariates and effects. Examples of effect types that can be accommodated include: (i) parametric effects of categorical covariates and linear or parametric effects of continuous covariates and their interactions, (ii) smooth nonlinear effects of continuous covariates, (iii)  bivariate smooth interactions, (iv) varying coefficient terms, and (v) random effects. For the specification of univariate and bivariate smooth functions we make use of Bayesian penalised B-splines \cite[P-splines,][]{Eilers1996, Lang2004}. That is, smooth functions are represented by (the tensor-product of) B-spline basis functions, and a random walk prior is placed on the basis coefficients to ensure smoothness. In addition, for the other effect types, appropriate matching prior distributions are chosen for the respective vectors of regression coefficients. This predictor formulation is the one employed in the so-called structured additive regression models \citep[STAR;][Chapter 9]{Fahrmeir2022}. We therefore term our approach as DDPstar. 

While previous works have explored integrating smooth effects into DP-based (or finite) mixture models using splines, this article expands upon this research in three key aspects: (i) unlike former contributions \cite[e.g.,][]{Inacio2013,Inacio2017,Inacio2022} we circumvent the critical and time-consuming task of selecting the optimal number (and location) of knots that define the spline basis by employing Bayesian P-splines (see more in Section \ref{sec:model}); (ii) in contrast to \cite{Chib2010} and \cite{Wiesenfarth14}, who model covariate effects through a smooth additive predictor (based on splines) and the error distribution as a DP mixture--thereby shifting the distribution of $y$ by the smooth predictor--we greatly enhance flexibility, while still retaining interpretability, by additionally mixing over the regression coefficients, thus allowing for potentially different smooth covariate effects in the mean of each mixture component; (iii) our approach permits complex and rich specifications for the components' means, accommodating multiple covariates and effect types. This differs from previous research, such as that by \cite{Berrettini2023}, which is constrained to using a single continuous covariate, thereby expanding the scope of possible models. All these advantages are attained while ensuring that all parameters of the DDPstar model have conjugate full conditional distributions, enabling straightforward Gibbs sampling and avoiding the need for specialised techniques or tuning of Metropolis--Hastings steps. Additionally, our method is implemented in the publicly available \texttt{R} package \texttt{DDPstar}, providing a user-friendly and broadly applicable approach.

The rest of the paper is structured as follows. Section \ref{sec:model} starts with the basics of the single-weights DDP mixture of normals model before we detail our approach based on a structured additive predictor for the mean of each normal component and how to conduct posterior inference. The performance of our method is validated in Section \ref{sim_main} using simulated data under a variety of challenging scenarios. In Section \ref{app_main} we apply our approach to data from toxicology, disease diagnosis, and agricultural studies. We conclude in Section \ref{sec:discussion} with a discussion. Details on posterior inference and extra results for both the simulation study and real data applications are available as Supplementary Materials.

\section{\large{\textsf{Dirichlet process mixture of normal structured additive regressions model}}}\label{sec:model}
Let $y \in \mathcal{Y} \subseteq \mathcal{R}$ be the univariate response variable and let  $ \boldsymbol{x}\in\mathcal{X}\subset \mathbb{R}^{p}$ be a $p-$dimensional covariate vector. Our starting point is the single-weights DDP mixture of normal distributions model \citep{deiorio2009}, which assumes that the conditional density function takes the following form
\begin{equation}\label{ddp}
p(y\mid\boldsymbol{x}) = \int \phi(y\mid\mu,\sigma^2)\text{d}G_{\boldsymbol{x}}(\mu,\sigma^2), \mbox{ where } 
G_{\boldsymbol{x}}(\cdot) = \sum_{l=1}^{\infty}\omega_l\delta_{(\mu_l(\boldsymbol{x}),\sigma_l^2)}(\cdot),
\end{equation}
with the weights $\{\omega_l\}_{l\geq 1}$ matching those from the standard stick-breaking representation in \eqref{sethu}. For simplicity, in what follows we consider a single continuous covariate and model its effect on the mean of each mixture component linearly
\begin{equation*}\label{eq:mean_lin}
\mu_{l}(x) = \beta_{l0} + \beta_{l1}x = \tilde{\boldsymbol{x}}^{\top}\boldsymbol{\beta}_l,\quad \tilde{\boldsymbol{x}} = (1,x)^{\top},\quad \boldsymbol{\beta}_l = (\beta_{l0},\beta_{l1})^{\top},\quad l\geq 1, 
\end{equation*}
which leads to a Dirichlet process mixture of normal linear (in the covariate's effect) models
\begin{align}\label{lddp}
p(y\mid\boldsymbol{x}) &= \int \phi(y\mid \tilde{\boldsymbol{x}}^{\top}\boldsymbol{\beta},\sigma^2)\text{d}G(\boldsymbol{\beta}, \sigma^2),\quad G\sim\text{DP}(\alpha, G_0(\boldsymbol{\beta},\sigma^2)) \nonumber\\
& = \sum_{l=1}^{\infty} \omega_l\phi(y\mid\tilde{\boldsymbol{x}}^{\top}\boldsymbol{\beta}_l,\sigma^2_l), \quad (\boldsymbol{\beta}_l,\sigma_l^2)\overset{\text{iid}}\sim G_0.
\end{align}
This model, by incorporating an infinite number of normal linear regression components, may seem very flexible. However, by assuming that the mixture weights are constant across $\boldsymbol{x}$, the conditional density and its functionals are greatly restricted. For instance, the mean regression structure is linear \cite[see, e.g.,][p.~165]{Dunson2007}. That is, this model is flexible in terms of non-Gaussian response, but not in terms of regression relationships. As a concrete example, Web Figure 1, first row, shows the fit of model \eqref{lddp} to data generated from a homoscedastic simple normal regression with a nonlinear trend. As can be observed, both the regression function and several profiles of the conditional density are poorly recovered. As noted in \citet[][p.~10]{Wade2023}, in single-weights DDPs, flexibility in the components-specific mean functions is key to achieve flexible, nonlinear mean regression structures. One possibility is to model the effect of continuous covariates within each mixture component through, for instance, B-spline basis functions \citep[][Chapter 8]{Fahrmeir2022}. In Web Figure 1, second row, we show the results of  fitting the model in \eqref{lddp} to the aforementioned simulated dataset but now considering a cubic B-spline basis expansion with ten interior knots. The model is now able to recover the true regression function and the profiles of the conditional density. However, it is widely acknowledged that the number (and location) of knots characterising the B-spline basis functions can significantly influence inferences. This is illustrated in the third and fourth rows of Web Figure 1, which display the results using a cubic B-spline basis expansion with no interior knots and forty interior knots, respectively. It is evident that employing no interior knots lacks the necessary flexibility, whereas the use of 40 interior knots leads to overfitting, especially in the case of the regression function and also to increased posterior uncertainty in the estimated density profiles. This example underscores the critical importance of selecting the `optimal' number of knots.

One possible approach is to use a model selection criterion. This strategy, successfully applied by \cite{Inacio2013,Inacio2017} and \cite{Inacio2022}, is most effective when dealing with a single continuous covariate. However, when faced with multiple continuous covariates, it becomes, in principle, necessary to fit the model in \eqref{lddp} for every conceivable combination of the number of interior knots, a task that may prove impractical. Another possibility would be to place a prior distribution on the number of knots, e.g., extending the approach of \cite{dimatteo2001}, but this would require reversible jump Markov chain Monte Carlo techniques, which tend to be challenging to implement efficiently in practice. 

In this work we favour the use of (Bayesian) penalised B-splines to model the effect of continuous covariates. P-splines have proven effective in estimating nonlinear effects, circumventing the sensitivity of unpenalised approaches to the number of knots (see last row of Web Figure 1) all while maintaining computational efficiency. The next sections are devoted to a detailed presentation of our proposal which is inspired by STAR models.

\subsection{\textsf{The DDPstar model}}\label{sec:DDPstar}
In its most general formulation, we write the mean of each mixture component in 
\eqref{ddp} as follows
\begin{equation}
\mu_{l}(\boldsymbol{x}) = \beta_{l0} + \sum_{a=1}^{A}f_{la}(\boldsymbol{v}_a),
\label{comp_add_pred}
\end{equation}
where $\boldsymbol{v}_a$ denotes subsets of the $p$ covariates in $\boldsymbol{x}$ and $f_{la}(\boldsymbol{v}_a)$ defines a generic representation of different types of functional  effects  depending on the covariate subset $\boldsymbol{v}_a$. As already noted, examples of effect types that can be accommodated into our framework include (i) parametric effects of categorical covariates and linear or parametric effects of continuous covariates and their interactions, (ii) smooth nonlinear effects of continuous covariates, (iii) bivariate smooth interactions or spatial eﬀects, (iv) varying coefficient terms, and (v) random effects \cite[see, e.g,][Chapter~9]{Fahrmeir2022}. For example, for a linear effect of a continuous covariate, we have $f_{la}(\boldsymbol{v}_a) = f_{l}^{\text{linear}}(v) = \beta_l v$, where $v$ is a univariate continuous covariate within the vector $\boldsymbol{x}$. In the case of nonlinear effects, $f_{la}(\boldsymbol{v}_a) = f_{l}^{\text{smooth}}(v)$, with $f_{l}^{\text{smooth}}$ being a smooth univariate function. 
Of particular interest here are cases (ii) and (iii), which we thoroughly explain in the next sections, whereas cases (iv) and (v) are detailed in Web Appendix B. For the sake of notational simplicity, we omit the specific function index ($a$) in the subsequent discussion.

\subsubsection{\textsf{Smooth nonlinear effects}}
We start with the case where $\boldsymbol{v}$ in \eqref{comp_add_pred} is constituted by a single continuous covariate, say $v$. To model smooth nonlinear effects, we consider the Bayesian analogue to P-splines \citep{Eilers1996} as introduced by \cite{Lang2004}. In particular, the smooth function is approximated by a linear combination of $J$ (cubic) B-spline basis functions on equidistant knots, i.e.,
\begin{equation}
f_{l}^{\text{smooth}}(v) = \sum_{j=1}^{J}\xi_{lj}B_{j}(v) = \mathbf{b}(v)^{\top}\boldsymbol{\xi}_{l},
\label{uni_smooth_fun}
\end{equation}
where $\mathbf{b}(v) = \left(B_{1}(v), \ldots, B_{J}(v)\right)^{\top}$ is the vector of B-spline basis functions evaluated at $v$ and $\boldsymbol{\xi}_{l} = (\xi_{l1},\ldots,\xi_{lJ})^{\top}$  is the vector of corresponding basis coefficients for the $l$th mixture component. P-splines rely on using a moderate to large number of basis functions, usually between 20 and 40, in combination with a penalty that enforces a smooth function estimation. Within a frequentist context, \cite{Eilers1996} proposed to penalise the squared $q$th order differences of adjacent basis coefficients. In the Bayesian framework, the $q$th order difference penalty is replaced by its stochastic analogue, i.e., a $q$th order random walk is used as a prior for the basis coefficients \citep{Lang2004}. The second-order random walk --the most popular in the literature and our choice here-- is defined by
\begin{equation}
\xi_{lj} = 2\xi_{l,j-1} - \xi_{l,j-2} + \epsilon_{lj},\quad \epsilon_{lj}\sim\text{N}(0,\tau_{l}^{2}), \quad l\geq 1,\quad j=3,\ldots,J.
\label{eq:random_walk}
\end{equation}
Usually, $\xi_{l1}$ and $\xi_{l2}$ are assigned noninformative priors, such that, $p(\xi_{l1})\propto \text{const}$ and $p(\xi_{l2})\propto \text{const}$. The random walk prior distribution variance, $\tau^{2}_{l}$, controls the amount of smoothing (it corresponds to the inverse of the smoothing parameter in the frequentist context), with small values corresponding to heavy smoothness and large values allowing considerable variation in the estimated smooth function. Note that we allow for a different amount of smoothing for each mixture component.

The second-order random walk prior in \eqref{eq:random_walk} induces the following joint Gaussian prior distribution for the vectors of basis coefficients $\boldsymbol{\xi}_{l}$ 
\begin{equation}
\label{eq:priorgamma}
p(\boldsymbol{\xi}_{l}\mid \tau_{l}^{2}) \propto |\mathbf{K}(\tau_{l}^{2})|_{+}^{1/2}\exp\left(-\frac{1}{2}\boldsymbol{\xi}_{l}^{\top}\mathbf{K}(\tau_{l}^{2})\boldsymbol{\xi}_{l}\right) \propto \left(\frac{1}{\tau_{l}^{2}}\right)^{\frac{\text{rank}(\mathbf{P})}{2}}\exp\left(-\frac{1}{2}\boldsymbol{\xi}_{l}^{\top}\mathbf{K}(\tau_{l}^{2})\boldsymbol{\xi}_{l}\right),
\end{equation}
where the precision matrix $\mathbf{K}(\tau_{l}^{2}) = \frac{1}{\tau_{l}^{2}}\mathbf{P}\in\mathbb{R}^{J\times J}$, with $\mathbf{P} = \mathbf{D}^{\top}\mathbf{D}$ and $\mathbf{D}$ being a second-order difference matrix, and $|\mathbf{A}|_{+}$ denotes the pseudo-determinant of the matrix $\mathbf{A}$ (i.e.,  the product of the non-zero eigenvalues). Because the penalty matrix $\mathbf{P}$ is rank deficient, $\mbox{rank} (\mathbf{P})=J-2$, the prior in \eqref{eq:priorgamma} is partially improper. This implies that there is a part of $f_{l}^{\text{smooth}}$ that is not penalised by the prior precision matrix.

To better understand the unpenalised part of $f_{l}^{\text{smooth}}$ as well as to make the mean function specification per component in \eqref{comp_add_pred} identifiable, $f_{l}^{\text{smooth}}$ is decomposed into two parts: a penalised and an unpenalised part. There are different ways to obtain such a decomposition; we follow \cite{Currie2006} and use the eigendecomposition of the penalty matrix $\mathbf{P}$. Let $\mathbf{P} = \mathbf{U}\mathbf{\Lambda}\mathbf{U}^{\top}$ be the eigendecomposition of $\mathbf{P}$, where $\mathbf{U}$ is the matrix of eigenvectors and $\mathbf{\Lambda}$ is the diagonal matrix of eigenvalues (with
eigenvalues arranged in order of increasing magnitude down the diagonal). Further denote by $\mathbf{U}_{+}$ ($\mathbf{\Lambda}_{+}$) and $\mathbf{U}_{0}$ ($\mathbf{\Lambda}_{0}$) the sub-matrices corresponding to the non-zero and zero eigenvalues, respectively, such that $\mathbf{U} = [\mathbf{U}_0, \mathbf{U}_{+}]$ and $\mathbf{\Lambda} = \text{blockdiag}\left(\mathbf{\Lambda}_{0}, \mathbf{\Lambda}_{+}\right)$. As noted before, for second-order random walk priors there are two zero eigenvalues ($\mbox{rank} (\mathbf{P})=J-2$). As such, $\mathbf{\Lambda}_{0}$ is a $2\times2$ matrix of zeroes, while $\mathbf{\Lambda}_{+}$ is a full-rank, $(J-2)\times(J-2)$, diagonal matrix. It is easy to show that \eqref{uni_smooth_fun} can then be reparameterised as
\begin{equation*}
f_{l}^{\text{smooth}}(v) = \mathbf{b}(v)^{\top}\boldsymbol{\xi}_{l} = \mathbf{b}(v)^{\top}\mathbf{U}\mathbf{U}^{\top}\boldsymbol{\xi}_{l} = \mathbf{x}(v)^{\top}\boldsymbol{\beta}_l + \mathbf{z}(v)^{\top}\boldsymbol{\gamma}_l,
\end{equation*}
where
\begin{equation*}
\mathbf{x}(v)^{\top} = \mathbf{b}(v)^{\top}\mathbf{U}_{0}, \quad \mathbf{z}(v)^{\top} = \mathbf{b}(v)^{\top}\mathbf{U}_{+}, \quad \boldsymbol{\beta}_l = \mathbf{U}^{\top}_{0}\boldsymbol{\xi}_{l}, \quad \boldsymbol{\gamma}_l = \mathbf{U}^{\top}_{+}\boldsymbol{\xi}_{l}.
\end{equation*}
We note that $\boldsymbol{\beta}_l^{\top}$ and $\boldsymbol{\gamma}_l^{\top}$ are vectors of length $2$ and $J-2$, respectively. Furthermore, $\boldsymbol{\xi}_{l} = \left[\mathbf{U}_{0},\mathbf{U}_{+}\right]\left(\boldsymbol{\beta}_l^{\top}, \boldsymbol{\gamma}_l^{\top}\right)^{\top}$, and thus the joint prior distribution in \eqref{eq:priorgamma} can be rewritten, in terms of the new vector of coefficients, $\left(\boldsymbol{\beta}_l^{\top}, \boldsymbol{\gamma}_l^{\top}\right)^{\top}$, as
\begin{equation}\label{eq:priorgamma_new}
p\left(\left(\boldsymbol{\beta}_l^{\top}, \boldsymbol{\gamma}_l^{\top}\right)^{\top}\;\middle|\;\tau_{l}^{2}\right) \propto \left(\frac{1}{\tau_{l}^{2}}\right)^{\frac{\text{rank}(\widetilde{\mathbf{P}})}{2}}\exp\left(-\frac{1}{2\tau_{l}^{2}}\left(\boldsymbol{\beta}_l^{\top}, \boldsymbol{\gamma}_l^{\top}\right)\mathbf{\Lambda}\left(\boldsymbol{\beta}_l^{\top}, \boldsymbol{\gamma}_l^{\top}\right)^{\top}\right), 
\end{equation}
In other words, the joint prior distribution in \eqref{eq:priorgamma_new} implies that $\boldsymbol{\beta}_l$ correspond to the unpenalised coefficients (with $p(\beta_{l1})\propto \text{const}$ and $p(\beta_{l2})\propto \text{const}$), while  $\boldsymbol{\gamma}_l$ is the vector of penalised coefficients with proper Gaussian prior distribution given by
\begin{equation}\label{eq:priorgamma_final}
p(\boldsymbol{\gamma}_l\mid \tau_{l}^{2}) \propto \left(\frac{1}{\tau_{l}^{2}}\right)^{\frac{J-2}{2}} \exp\left(-\frac{1}{2\tau_{l}^{2}}\boldsymbol{\gamma}_l^{\top}\mathbf{\Lambda}_{+}\boldsymbol{\gamma}_l\right).
\end{equation}
Note that although based on the eigendecomposition we have that $\mathbf{x}(v)^{\top} = \mathbf{b}_{l}(v)^{\top}\mathbf{U}_{0}$, this is equivalent to consider $\mathbf{x}(v)^{\top} = (1,v)$  \cite[see, e.g.,][]{Lee2010}. As such, in P-splines with second-order random walk priors, the space of functions that are not penalised corresponds to the polynomials of degree $1$. This implies that when $\tau_{l}^{2}\rightarrow 0$, the estimated function approaches a linear effect in that component. Moreover, this reparametrisation makes clear that the B-spline basis expansion of $f_{l}^{\text{smooth}}$ in \eqref{uni_smooth_fun} includes an intercept (constant term). Given that there is already an intercept in the model for the mean of each component (see Equation \eqref{comp_add_pred}), when constructing univariate smooth functions using P-splines, the intercept is removed to avoid identifiability issues (i.e., we consider $\mathbf{x}(v)^{\top} = v$). Note that this approach also permits the incorporation of univariate smooth effects for multiple continuous covariates in \eqref{comp_add_pred}, effectively circumventing identifiability issues.
\subsubsection{\textsf{Bivariate smooth surfaces}}
We now move onto the case where $\boldsymbol{v}$ in \eqref{comp_add_pred} is constituted by two continuous covariates, say $v_1$ and $v_2$, and we are interested in modelling a smooth bivariate surface jointly defined over $v_1$ and $v_2$. In the case of a spatial effect, $v_1$ and $v_2$ typically represent coordinate information about the spatial location.

When extending the principles of P-splines to the bivariate case, one first approximates the smooth bivariate surface using the tensor-product of two marginal B-splines bases, i.e., 
\begin{equation*}
f_{l}^{\text{bivariate}}(v_1, v_2) = \sum_{j_1 = 1}^{J_1}\sum_{j_2 = 1}^{J_2}\xi_{lj_1j_2}B_{1j_1}(v_1)B_{2j_2}(v_2) = \left(\mathbf{b}_{1}(v_1)^{\top}\otimes\mathbf{b}_{2}(v_2)^{\top}\right)\boldsymbol{\xi}_{l},
\end{equation*}
where $\mathbf{b}_{1}(v_1)$ and $\mathbf{b}_{2}(v_2)$ are the vectors containing the B-spline basis functions evaluations, $\otimes$ denotes the Kronecker product, and $\boldsymbol{\xi}_{l} = \left(\xi_{l11}, \ldots, \xi_{l1J_2}, \ldots, \xi_{lJ_11}, \ldots, \xi_{lJ_1J_2}\right)^{\top}$ is the vector of coefficients. Smoothness is achieved by penalising (the sum of squares of) second-order coefficient differences along $v_1$ and $v_2$ \cite[for details, see,][]{Eilers2003}, which translates into the following partially improper Gaussian prior distribution
\begin{equation}\label{eq:priorgamma_2D}
p(\boldsymbol{\xi}_{l}\mid \tilde{\tau}_{l1}^{2}, \tilde{\tau}_{l2}^{2}) \propto |\mathbf{K}(\tilde{\tau}_{l1}^{2}, \tilde{\tau}_{l2}^{2})|_{+}^{1/2} \exp\left(-\frac{1}{2}\boldsymbol{\xi}_{l}^{\top}\mathbf{K}(\tilde{\tau}_{l1}^{2}, \tilde{\tau}_{l2}^{2})\boldsymbol{\xi}_{l}\right),
\end{equation}
where
\begin{equation*}
\mathbf{K}(\tilde{\tau}_{l1}^{2}, \tilde{\tau}_{l2}^{2}) = \frac{1}{\tilde{\tau}_{l1}^2}\left(\mathbf{P}_{1}\otimes\boldsymbol{I}_{J_2}\right) + \frac{1}{\tilde{\tau}_{l2}^2}\left(\boldsymbol{I}_{J_1}\otimes \mathbf{P}_{2}\right),
\end{equation*}
with $\mathbf{P}_{1} = \mathbf{D}_{1}^{\top}\mathbf{D}_{1}$ and $\mathbf{P}_{2} = \mathbf{D}_{2}^{\top}\mathbf{D}_{2}$. We note that by using two prior variances, $\tilde{\tau}_{l1}^{2}$ and $\tilde{\tau}_{l2}^{2}$, the prior distribution in \eqref{eq:priorgamma_2D} permits a different amount of smoothing for $v_1$ and $v_2$. 

As for the univariate case, the penalty matrix $\mathbf{P} = \left(\mathbf{P}_{1}\otimes\boldsymbol{I}_{J_2}\right) + \left(\boldsymbol{I}_{J_1}\otimes \mathbf{P}_{2}\right)$ is rank deficient, with $\text{rank}(\mathbf{P})=J_1J_2-4$. We proceed similarly and decompose the tensor-product P-spline smooth bivariate surface into a penalised and an unpenalised part using the eigendecomposition of the marginal penalties $\mathbf{P}_1 = \mathbf{U}_{1}\mathbf{\Upsilon}_{1}\mathbf{U}_{1}^{\top}$ and  $\mathbf{P}_{2} = \mathbf{U}_{2}\mathbf{\Upsilon}_{2}\mathbf{U}_{2}^{\top}$. It can be shown that \cite[for details, see][]{Lee2010}
\begin{equation*}
\begin{split}
f_{l}^{\text{bivariate}}(v_1, v_2) & = \left(\mathbf{b}_{1}(v_1)^{\top}\otimes\mathbf{b}_{2}(v_2)^{\top}\right)\boldsymbol{\xi}_{l}\\ & = \left(\mathbf{x}_{1}(v_1)^{\top} \otimes \mathbf{x}_{2}(v_2)^{\top}\right)\boldsymbol{\beta}_l + \left(\mathbf{u}_{1}(v_1)^{\top}\otimes \mathbf{x}_{2}(v_2)^{\top}, \mathbf{x}_{1}(v_1)^{\top}\otimes \mathbf{u}_{2}(v_2)^{\top}, \mathbf{u}_{1}(v_1)^{\top}\otimes \mathbf{u}_{2}(v_2)^{\top}\right)\boldsymbol{\gamma}_l\\
& \equiv \mathbf{x}(v_1, v_2)^{\top}\boldsymbol{\beta}_l + \mathbf{u}(v_1, v_2)^{\top}\boldsymbol{\gamma}_l,
\end{split}
\end{equation*}
where the symbol $\equiv$ is used to indicate that the design vectors in the second and third rows have the same elements but in a different order,
\begin{equation}
\mathbf{x}_{1}(v_{1})^{\top} = (1, v_1), \quad\mathbf{u}_{1}(v_1)^{\top} = \mathbf{b}_{1}(v_{1})^{\top}\mathbf{U}_{1+}, \quad
\mathbf{x}_{2}(v_{2})^{\top} = (1, v_2), \quad\mathbf{u}_{2}(v_2)^{\top} = \mathbf{b}_{2}(v_{2})^{\top} \mathbf{U}_{2+},
\label{eq:dm_2d}
\end{equation}
and 
\begin{align}
\mathbf{x}(v_1, v_2)^{\top} &  = (1, v_1, v_2, v_1v_2), \nonumber\\
\mathbf{u}(v_1, v_2)^{\top} &  = \left(\mathbf{u}_{1}(v_1)^{\top}, \mathbf{u}_{2}(v_2)^{\top}, v_2\mathbf{u}_{1}(v_1)^{\top}, v_1\mathbf{u}_{2}(v_2)^{\top}, \mathbf{u}_{1}(v_1)^{\top}\otimes\mathbf{u}_{2}(v_2)^{\top}\right)\label{eq:dm_random_2d}.
\end{align}
In this case, $\boldsymbol{\beta}_l^{\top}$ and $\boldsymbol{\gamma}_l^{\top}$ are vectors of length $4$ and $J_1J_2-4$, respectively, with $\boldsymbol{\beta}_l$ corresponding to the unpenalised coefficients (with $p(\beta_{lk})\propto \text{const}$, $k =1,\ldots, 4$), and  $\boldsymbol{\gamma}_l$ to the penalised coefficients, with proper Gaussian prior distribution given by
\begin{equation}\label{eq:priorgamma_2D_final}
p(\boldsymbol{\gamma}_{l}\mid \tilde{\tau}_{l1}^{2}, \tilde{\tau}_{l2}^{2}) \propto |\widetilde{\mathbf{K}}(\tilde{\tau}_{l1}^{2}, \tilde{\tau}_{l2}^{2})|^{1/2} \exp\left(-\frac{1}{2}\boldsymbol{\gamma}_{l}^{\top}\widetilde{\mathbf{K}}(\tilde{\tau}_{l1}^{2}, \tilde{\tau}_{l2}^{2})\boldsymbol{\gamma}_{l}\right),
\end{equation}
where the precision matrix is
\begin{equation}
\widetilde{\mathbf{K}}(\tilde{\tau}_{l1}^{2}, \tilde{\tau}_{l2}^{2}) = \text{blockdiag}\left(\frac{1}{\tilde{\tau}_{l1}^2}\mathbf{\Upsilon}_{1+}, \frac{1}{\tilde{\tau}_{l2}^2}\mathbf{\Upsilon}_{2+}, \frac{1}{\tilde{\tau}_{l1}^2}\mathbf{\Upsilon}_{1+}, \frac{1}{\tilde{\tau}_{l2}^2}\mathbf{\Upsilon}_{2+}, \frac{1}{\tilde{\tau}_{l1}^2}\mathbf{\Upsilon}_{1+} \otimes \boldsymbol{I}_{J_2 - 2} + \frac{1}{\tilde{\tau}_{l2}^2}\boldsymbol{I}_{J_1 - 2}\otimes \mathbf{\Upsilon}_{2+}\right).
\label{eq:2d_prec_mat}
\end{equation}
A technical note is in order here. For notational convenience and simplicity, in \eqref{eq:dm_2d} we have considered $\mathbf{x}_{d}(v_{d})^{\top} = (1, v_d)$, $d = 1,2$. However, to ensure that the precision matrix associated with $\boldsymbol{\gamma}_l$  matches the one presented in \eqref{eq:2d_prec_mat}, certain adjustments are required. One possibility is to substitute $v_d$ with a centred version based on the covariate observations. Another, more comprehensive method is discussed in \citet[p.~ 346]{wood2013}, and this is the one we have used in our implementation.

Note that $\mathbf{u}(v_1, v_2)^{\top}$ in \eqref{eq:dm_random_2d}, related to the penalised part of the tensor-product P-spline smooth bivariate surface, is composed by $5$ building blocks (subvectors), and so $\boldsymbol{\gamma}_{l}^{\top}$ can be seen as the concatenation of five subvectors of penalised coefficients, i.e., $\boldsymbol{\gamma}_{l}^{\top} = \left(\boldsymbol{\gamma}_{l1}^{\top}, \boldsymbol{\gamma}_{l2}^{\top}, \boldsymbol{\gamma}_{l3}^{\top}, \boldsymbol{\gamma}_{l4}^{\top}, \boldsymbol{\gamma}_{l5}^{\top}\right)$. In fact, each block in the precision matrix \eqref{eq:2d_prec_mat} corresponds to one of these coefficients subvectors. Moreover, the block structure of $\mathbf{u}(v_1, v_2)^{\top}$ also leads to an interesting ANOVA-type decomposition of the penalised part of the bivariate surface in five different smooth terms
\begin{equation}
\begin{split}
f_{l}^{\text{bivariate}}(v_1, v_2)  & = \beta_{l1} + \beta_{l2}v_1 + \beta_{l3}v_2 + \beta_{l4}v_1v_2 \\ & 
+ \underbrace{f_{l1}(v_1)}_{\mathbf{u}_{1}(v_1)^{\top}\boldsymbol{\gamma}_{l1}} + \underbrace{f_{l2}(v_2)}_{\mathbf{u}_{2}(v_2)^{\top}\boldsymbol{\gamma}_{l2}} + \underbrace{v_2h_{l1}(v_1)}_{v_2\mathbf{u}_{1}(v_1)^{\top}\boldsymbol{\gamma}_{l3}} + \underbrace{v_1h_{l2}(v_2)}_{v_1\mathbf{u}_{2}(v_2)^{\top}\boldsymbol{\gamma}_{l4}} +  \underbrace{f_{l(1,2)}(v_1, v_2)}_{\left(\mathbf{u}_{1}(v_1)^{\top}\otimes\mathbf{u}_{2}(v_2)^{\top}\right)\boldsymbol{\gamma}_{l5}}.
\end{split}
\label{eq:2d_decomp}
\end{equation}
In other words, there are two main \textit{pure} smooth effects along $v_1$ and $v_2$, $f_{l1}(v_1)$ and $f_{l2}(v_2)$, two varying-coefficient terms, $v_2h_{l1}(v_1)$ and $v_1h_{l2}(v_2)$, and a \textit{pure} nonlinear-by-nonlinear interaction term, $f_{l(1,2)}(v_1, v_2)$. For further details and insights we refer the reader to \cite{Lee2013} and \cite{rod_alv_17}.

Examining the precision matrix \eqref{eq:2d_prec_mat} reveals that, despite the five smooth terms, only two prior variances control their smoothness, $\tilde{\tau}^2_{l1}$ and $\tilde{\tau}^2_{l2}$, and the same prior variances apply to both main effects and interaction terms. Moreover, the last block in \eqref{eq:2d_prec_mat} depends on both prior variances. As discussed in \cite{Kneib2019} and \cite{Bach2022}, this has important implications for posterior simulation. The full conditional posterior distribution of $\tilde{\tau}^2_{l1}$ and $\tilde{\tau}^2_{l2}$ involves the determinant $|\widetilde{\mathbf{K}}(\tilde{\tau}_{l1}^{2}, \tilde{\tau}_{l2}^{2})|$, and this precludes the use of Gibbs sampling for posterior simulation. To circumvent this problem and enable posterior simulation through Gibbs sampling, we extend here the so-called PS-ANOVA model proposed by \cite{Lee2013} to the Bayesian setting (BPS-ANOVA). The idea underlying BPS-ANOVA is to use a different prior variance for each smooth term in \eqref{eq:2d_decomp}, i.e., each block in the prior precision matrix for $\boldsymbol{\gamma}_{l}$, see Equation \eqref{eq:2d_prec_mat}, will have its own prior variance. That is, under the BPS-ANOVA approach, the penalised part of the tensor-product P-spline smooth bivariate surface is represented as the sum of five sets of mutually independent coefficients $\boldsymbol{\gamma}_{lk}$ ($k = 1, \ldots, 5$), each with a proper Gaussian prior distribution and a different prior variance, say ${\tau}_{lk}^2$ ($k = 1, \ldots, 5$). To make things concrete, we first need to introduce further notation. Let
\begin{equation*}
\begin{alignedat}{2}
\mathbf{z}_{1}(\boldsymbol{v}) = \mathbf{u}_1(v_1), \quad \mathbf{z}_{2}(\boldsymbol{v}) = \mathbf{u}_2(v_2),  & \quad
\mathbf{z}_{3}(\boldsymbol{v}) = v_2\mathbf{u}_1(v_1), \quad \mathbf{z}_{4}(\boldsymbol{v}) = v_1\mathbf{u}_2(v_2),
&\quad&\raisebox{-.5\normalbaselineskip}[0pt][0pt]{(see Eq. \eqref{eq:dm_random_2d} and \eqref{eq:2d_decomp}),} \\
\mathbf{z}_{5}(\boldsymbol{v}) & = \mathbf{u}_1(v_1)\otimes\mathbf{u}_2(v_2),
\end{alignedat}
\end{equation*}
and 
\begin{equation*}
\mathbf{\Lambda}_{1+} = \mathbf{\Lambda}_{3+} = \mathbf{\Upsilon}_{1+},\quad \mathbf{\Lambda}_{2+} = \mathbf{\Lambda}_{4+} = \mathbf{\Upsilon}_{2+},\quad \mathbf{\Lambda}_{5+} = \mathbf{\Upsilon}_{1+} \otimes \boldsymbol{I}_{J_2 - 2} + \boldsymbol{I}_{J_1 - 2}\otimes  \mathbf{\Upsilon}_{2+} \quad \text{(see Eq. \eqref{eq:2d_prec_mat}).}
\end{equation*}
Therefore, the tensor-product P-spline smooth bivariate surface in Equation \eqref{eq:2d_decomp} can be expressed, excluding the intercept, as
\begin{equation*}
f_{l}^{\text{bivariate}}(v_1, v_2) = \beta_{l2}v_1 + \beta_{l3}v_2 + \beta_{l4}v_1v_2 + \sum_{k=1}^{5}\mathbf{z}_{k}\left(\boldsymbol{v}\right)^{\top}\boldsymbol{\gamma}_{lk},
\end{equation*}
and, under the BPS-ANOVA model, the prior Gaussian distributions for $\boldsymbol{\gamma}_{lk}$, $k = 1, \ldots, 5$, are given by
\begin{equation}\label{eq:priorgamma_2d_final}
p(\boldsymbol{\gamma}_{lk}\mid 
\tau_{lk}^{2}) \propto \left(\frac{1}{
\tau_{lk}^{2}}\right)^{\frac{\text{rank}\left(\mathbf{\Lambda}_{k+}\right)}{2}}\exp\left(-\frac{1}{2\tau_{lk}^{2}}\boldsymbol{\gamma}_{lk}^{\top}\mathbf{\Lambda}_{k+}\boldsymbol{\gamma}_{lk}\right).
\end{equation}

\subsection{Prior specification and posterior inference}\label{sec:prior_post}
Using the elements introduced in Section \ref{sec:DDPstar}, we can compactly write the mean of each normal component as
\begin{equation}
\mu_{l}(\boldsymbol{x}) = \underbrace{\mathbf{x}^{\top}\boldsymbol{\beta}_{l}}_{\mbox{Parametric}} + \underbrace{\sum_{r = 1}^{R}\mathbf{z}_{r}(\boldsymbol{v}_r)^{\top}\boldsymbol{\gamma}_{lr}}_{\mbox{Smooth}}.
\label{eq:DDPstar_compact}
\end{equation}
Here $\mathbf{x}^{\top}\boldsymbol{\beta}_{l}$ contains all parametric effects, including, by default, the intercept, parametric effects of categorical covariates and linear or parametric effects of continuous covariates and their interactions, as well as, the unpenalised terms associated with the P-splines formulation, whereas $\sum_{r = 1}^{R}\mathbf{z}_{r}(\boldsymbol{v}_r)^{\top}\boldsymbol{\gamma}_{lr}$ contains the P-splines' penalised terms. Our model for the conditional density can therefore be written as
\begin{equation}\label{final_model}
p(y\mid \boldsymbol{x}) =
\int \phi(y\mid \mathbf{x}^{\top}\boldsymbol{\beta} + \sum_{r = 1}^{R}\mathbf{z}_{r}(\boldsymbol{v}_r)^{\top}\boldsymbol{\gamma}_{r},\sigma^2)\text{d}G(\boldsymbol{\beta},\boldsymbol{\gamma}_{1}, \ldots, \boldsymbol{\gamma}_{R},\sigma^2),\quad
G(\cdot)= \sum_{l=1}^{\infty}\omega_l\delta_{(\boldsymbol{\beta}_l, \boldsymbol{\gamma}_{l1},\ldots, \boldsymbol{\gamma}_{lR},\sigma_l^2)}(\cdot),
\end{equation}
where the weights $\{\omega_l\}_{l\geq 1}$ match those from the standard stick-breaking representation in \eqref{sethu}.
The model is completed with the prior specification and we specify a conditionally conjugate centring distribution
\begin{equation*}
\boldsymbol{\beta}_l \mid  \mathbf{m}_{\boldsymbol{\beta}}, \mathbf{H}_{\boldsymbol{\beta}}\overset{\text{iid}}\sim\text{N}(\mathbf{m}_{\boldsymbol{\beta}},\mathbf{H}_{\boldsymbol{\beta}}), \quad \boldsymbol{\gamma}_{lr}\mid\tau_{lr}^2\overset{\text{ind}}\sim\text{N}(\mathbf{0},\tau_{lr}^2\mathbf{K}_r^{-1}),\quad \sigma_l^2\overset{\text{iid}}\sim\text{IG}(a_{\sigma^2},b_{\sigma^2}),
\end{equation*}
with conjugate hyperpriors  
\begin{equation*}
(\mathbf{m}_{\boldsymbol{\beta}},\mathbf{H}_{\boldsymbol{\beta}}^{-1})\sim \text{N}(\mathbf{m}_0,\mathbf{H}_0)\times \text{W}(\nu,(\nu\Psi)^{-1}),
\quad \tau_{lr}^{2}\overset{\text{iid}}\sim\text{IG}(a_{\tau^2}, b_{\tau^2}),
\end{equation*}
where $\text{IG}(a,b)$ denotes an inverse gamma distribution with shape parameter $a$ and scale parameter $b$ and $\text{W}(\nu,(\nu\Psi)^{-1})$ denotes a Wishart distribution with $\nu$ degrees of freedom and expectation $\Psi^{-1}$. The specific forms of $\mathbf{K}_{r}$ in the prior distribution for $\boldsymbol{\gamma}_{lr}$ can be seen in \eqref{eq:priorgamma_final} and \eqref{eq:priorgamma_2d_final} (and in Equations (B2)--(B4) of Web Appendix B for the case of varying coefficient terms and random effects). The precision parameter $\alpha$ (see Equation \eqref{sethu}) can either be fixed or have a prior distribution placed on it. In this work, we follow the latter approach and, for conjugacy reasons, let $\alpha\sim\text{Gamma}(a_{\alpha}, b_{\alpha})$.

For posterior inference, we use the blocked Gibbs sampler of \cite{Ishwaran2001}, which relies on truncating the stick-breaking representation to a finite number of components, say $L$. Hence the mixing distribution $G$ in \eqref{final_model} is replaced by $G^{L}(\cdot) = \sum_{l=1}^{L}\omega_l\delta_{(\boldsymbol{\beta}_l, \boldsymbol{\gamma}_{l1},\ldots,\boldsymbol{\gamma}_{lR},\sigma_l^2)}(\cdot)$, with $\eta_L=1$ to ensure that the weights sum up to one. We shall emphasise that $L$ is not the exact number of mixture components expected to be observed, but rather an upper bound on it, as some of the components may not be occupied. We further note that there is an interplay between $L$ and the precision parameter $\alpha$. The value of $L$ can be chosen considering distributional properties for the tail of the stick-breaking weights, $U_L = \sum_{l= L+1}^{\infty}\omega_l$; \cite{Ishwaran2000} have shown that $E(U_L)=\{\alpha/(\alpha+1)\}^L$. Furthermore, denoting by $L^{*}(\leq L \leq n)$ the number of occupied mixture components, we have that $E(L^{*}\mid \alpha)\approx \alpha \log((\alpha + n)/\alpha)$ and $\text{var}(L^{*}\mid \alpha) \approx \alpha\{\log((\alpha + n)/\alpha) - 1\}$ 
\citep[see, e.g.,][]{Liu1996}. This information can be useful to determine an appropriate value of $L$. For instance, it may be reasonable to set $L>E(L^{*}\mid \alpha) + 2\sqrt{\text{var}(L^{*}\mid \alpha)}$. In the posterior samples it is also possible to check the number of occupied components and if that is close to or equal to $L$ for most of the iterations, the analysis should be redone with a larger $L$. We favour the use of the blocked Gibbs sampler, over other posterior sampling algorithms, due to its ease of implementation and because it allows for full posterior inference for general regression functionals.

Upon the introduction of latent variables that identify the mixture component to which each observation belongs to, the model for the data can be written hierarchically. From this hierarchical representation, it is straightforward to derive the full conditional distributions of all model parameters, which are available in closed form, thus allowing for ready posterior simulation through Gibbs sampling. All details are available in Web Appendix C. Each posterior sample for $\{(\omega_l,\boldsymbol{\beta}_l,\boldsymbol{\gamma}_{l1},\ldots, \boldsymbol{\gamma}_{lR},\sigma_l^2)\}_{l=1}^{L}$ provides a posterior realisation for $G^L$ directly through its definition and therefore for any point of interest, say $(y_0,\mathbf{x}_0)$, we can obtain posterior realisations of the predictive conditional density, $p(y_0\mid \mathbf{x}_0)$. Moreover, posterior realisations of the predictive conditional mean, variance, and quantiles can also be easily obtained (detailed expression are given in Web Appendix D).

\section{\large{\textsf{Simulation study}}}\label{sim_main}
This section reports the results of a simulation study conducted to evaluate the empirical performance of the proposed model, DDPstar, for conducting inference about different functionals of the conditional response distribution, namely the conditional densities, which are our key inferential objects, the conditional mean and variance, and the three conditional quartiles. We do not include comparisons against  other methods as our goal is not to claim superior performance of DDPstar over competing covariate-dependent Dirichlet process mixture models. Instead, we restate that our aim is to provide a tractable density regression model that performs well in a variety of scenarios, whose performance can be readily assessed through model fitting checks, and for which posterior inference is straightforward to implement.
Simulations are performed in the \texttt{R} environment \citep{R23}, using the \texttt{R} package \texttt{DDPstar} that accompanies this paper. Plots are generated using the \texttt{R} package \texttt{ggplot2} \citep{Wickham16}.

\subsection{\textsf{Simulation scenarios and implementation details}}\label{scenarios}
We explore a broad spectrum of simulation scenarios, encompassing situations involving a single continuous covariate as well as two continuous covariates. In the latter case, we investigate both additive and interaction structures. For each scenario, $100$ datasets are generated for sample sizes $n \in \{300, 500, 1000\}$. The scenarios we examine are as follows.
\begin{itemize}
\item \textbf{Scenario I}. Heteroscedastic nonlinear normal regression
$$y_i\mid x_i\overset{\text{ind.}}\sim\text{N}\left(x_i-4x_i^3,(1+x_i)^2\right), \quad x_i\overset{\text{iid}}\sim\text{U}(0,1).$$

\item \textbf{Scenario II}. Mixture of nonlinear non-normal regressions
$$y_i\mid x_i\overset{\text{ind.}}\sim0.5\,\text{SN}(\xi =x_i^2,\omega=0.25,\alpha=2)+0.5\,\text{t}(\mu=\sin(\pi x_i),\sigma=0.25,\nu=5), \quad x_i\overset{\text{iid}}\sim\text{U}(0,1),$$
where $\text{SN}(\xi,\omega,\alpha)$ denotes a skew normal distribution with location $\xi$, scale $\omega$, and skewness parameter $\alpha$, and $\text{t}(\mu,\sigma,\nu)$ denotes a (scaled and shifted) $t$ distribution with mean $\mu$, scale parameter $\sigma$, and degrees of freedom $\nu$.
\item \textbf{Scenario III}. Covariate-dependent mixture of normal homoscedastic regressions
$$y_i\mid x_i\overset{\text{ind.}}\sim\exp(-2x_i)\,\text{N}(x_i,0.01)+\{1-\exp(-2x_i)\}\,\text{N}(x_i^4,0.04),\quad x_i\overset{\text{iid}}\sim\text{U}(0,1).$$
This scenario has been considered in previous studies, including \cite{Dunson2007} and \cite{Dunson2008}, and it poses a challenge for our model due to our assumption of covariate-independent weights.
\item \textbf{Scenario IV}. Nonstationary mean function and covariate-dependent variance
\begin{align*}
&y_i\mid x_i \overset{\text{ind.}}\sim\text{N}(m(x_i), \sigma^2(x_i)),\\
&m(x_i) = 
\begin{cases} 
0,  & \text{if } -2\leq x_i \leq 2,\\
2x_i-4, & \text{if } 2 <x_i\leq 5,\\
6,& \text{if } 5 < x_i  \leq 10,
\end{cases} 
\quad 
\sigma^2(x_i) = 
\begin{cases}
0.2^2, & \text{if } -2\leq x_i \leq 2,\\
0.05^2, & \text{if } 2 <x_i\leq 5,\\
(x_i-5)^2/15 + 0.01,& \text{if } 5 < x_i  \leq 10,\\
\end{cases}
\end{align*}
and $x_i\overset{\text{iid}}\sim \text{U}(-2,10)$. This scenario has been used by \cite{Wade2023} to illustrate drawbacks of single-weights DDPs.
\item \textbf{Scenario V}. Mixture of nonlinear normal regressions (two continuous covariates)
$$y_i\mid x_{1i}, x_{2i}\overset{\text{ind.}} \sim 0.5\,\text{N}(g_{11}(x_{1i}) + g_{12}(x_{2i}), 0.25^2) + 0.5\,\text{N}(\mu= g_{21}(x_{1i}) + g_{22}(x_{2i}), 0.25^2).$$
\item \textbf{Scenario VI}. Mixture of nonlinear non-normal regressions (two continuous covariates)
$$y_i\mid x_{1i}, x_{2i}\overset{\text{ind.}} \sim 0.5\,\text{SN}(\xi = g_{11}(x_{1i}) + g_{12}(x_{2i}),\omega=0.25,\alpha=2) + 0.5\,\text{t}(\mu= g_{21}(x_{1i}) + g_{22}(x_{2i}),\sigma=0.25,\nu=5).$$
\end{itemize}
For Scenarios V and VI, we consider $x_{1i}, x_{2i}\overset{\text{iid}}\sim\text{U}(0,1)$, $g_{11}(x_{1}) = (x_1 - 0.5)^2$, $g_{12}(x_{2}) = (x_2 - 0.5)^2$, $g_{21}(x_{1}) = \exp(x_1)\sin\left(13(x_1-0.6)^2\right)$, and $g_{22}(x_{2}) = \exp(-x_2)\sin(7x_2)$. Note that these two scenarios correspond to the case of two continuous covariates and an additive structure.
\begin{itemize}
\item \textbf{Scenario VII}. Mixture of nonlinear normal regressions (interaction between two continuous covariates)
$$y_i\mid x_{1i}, x_{2i}\overset{\text{ind.}} \sim 0.5\,\text{N}(g_1(x_{1i}, x_{2i}), 0.25^2) + 0.5\,\text{N}(g_2(x_{1i}, x_{2i}),0.25^2).$$
\item \textbf{Scenario VIII}. Mixture of nonlinear non-normal regressions (interaction between two continuous covariates)
$$y_i\mid x_{1i}, x_{2i}\overset{\text{ind.}} \sim 0.5\,\text{SN}(\xi = g_1(x_{1i}, x_{2i}),\omega=0.25,\alpha=2) + 0.5\,\text{t}(\mu= g_2(x_{1i}, x_{2i}),\sigma=0.25,\nu=5).$$
\end{itemize}
For Scenarios VII and VIII, we consider $x_{1i}, x_{2i}\overset{\text{iid}}\sim\text{U}(0,1)$, $g_1(x_{1}, x_{2}) = \cos\left(2\pi\sqrt{(x_1 - 0.5)^2 + (x_2 - 0.5)^2)}\right)$, and $g_2(x_{1}, x_{2}) = 1.9\exp(x_1)\sin\left(13(x_1-0.6)^2\right)\exp(-x_2)\sin(7x_2)$. Both scenarios correspond to a situation involving two continuous covariates and an interaction structure.

For scenarios involving a single continuous covariate, Scenario I to IV, the mean of each component is specified as $\mu_{l}(x) = f_l(x)$, and $f_l$ is approximated using a cubic B-spline basis of dimension $J = 23$. For Scenarios V and VI, we consider $\mu_{l}(x_1, x_2) = f_{l1}(x_1) + f_{l2}(x_2)$, and a cubic B-spline basis of dimension $J = 23$, for both $f_{l1}$ and $f_{l2}$, is used again. For Scenarios VII and VIII we consider $\mu_{l}(x_1, x_2) = f_{l}(x_1, x_2)$, and represent the bivariate function using the tensor product of two marginal cubic B-splines bases of dimension $J_1 = 23$ and $J_2=23$. In
all cases, we use second-order random walks.

To facilitate prior specification, we standardise responses and covariates, and use the prior distributions outlined in Section \ref{sec:prior_post}. Following \cite{Lang2004}, for the penalised or nonlinear part of our P-splines specification, we set $a_{\tau^2}=1$ and $b_{\tau^2}=0.05$, which is a common default. For $\boldsymbol{\beta}_l$, the vector containing the coefficients associated to the unpenalised or linear terms, we use $\mathbf{m}_0=\mathbf{0}_Q$, $\mathbf{H}_0=10I_{Q}$, $\nu=Q+2$, and $\Psi=I_{Q}$, where $Q$ is the length of vector $\boldsymbol{\beta}_l$ (including the intercept). Regarding the prior for $\sigma_l^2$, we choose $a_{\sigma^2}=2$ and $b_{\sigma^2}=0.5$. Selecting $a_{\sigma^2}=2$ results in a prior distribution with infinite variance centred around a finite mean ($b_{\sigma^2}=0.5$), thereby favouring within-component variances smaller than one. Given that responses are standardised, and thus have a marginal variance of one, it is reasonable to expect the within-component variances to be smaller than the marginal variance. We further set $a_{\alpha}=b_{\alpha}=2$ and $L = 20$, which results in a truncation error of $E(U_{20}\mid \alpha)\approx 0.0002$. These choices lead, for the largest sample size of $n=1000$, to an expected (standard deviation) number of occupied components of $7 (3)$, and thus setting $L=20$ seems appropriate a priori. Lastly, for each scenario and simulated dataset, posterior inference is based  on $4000$ samples after a burn-in period of $1000$ iterations of the Gibbs sampler is discarded.

\subsection{\textsf{Results}}
\begin{figure}[H]
\begin{center}
\includegraphics[scale = 0.37, page = 1]{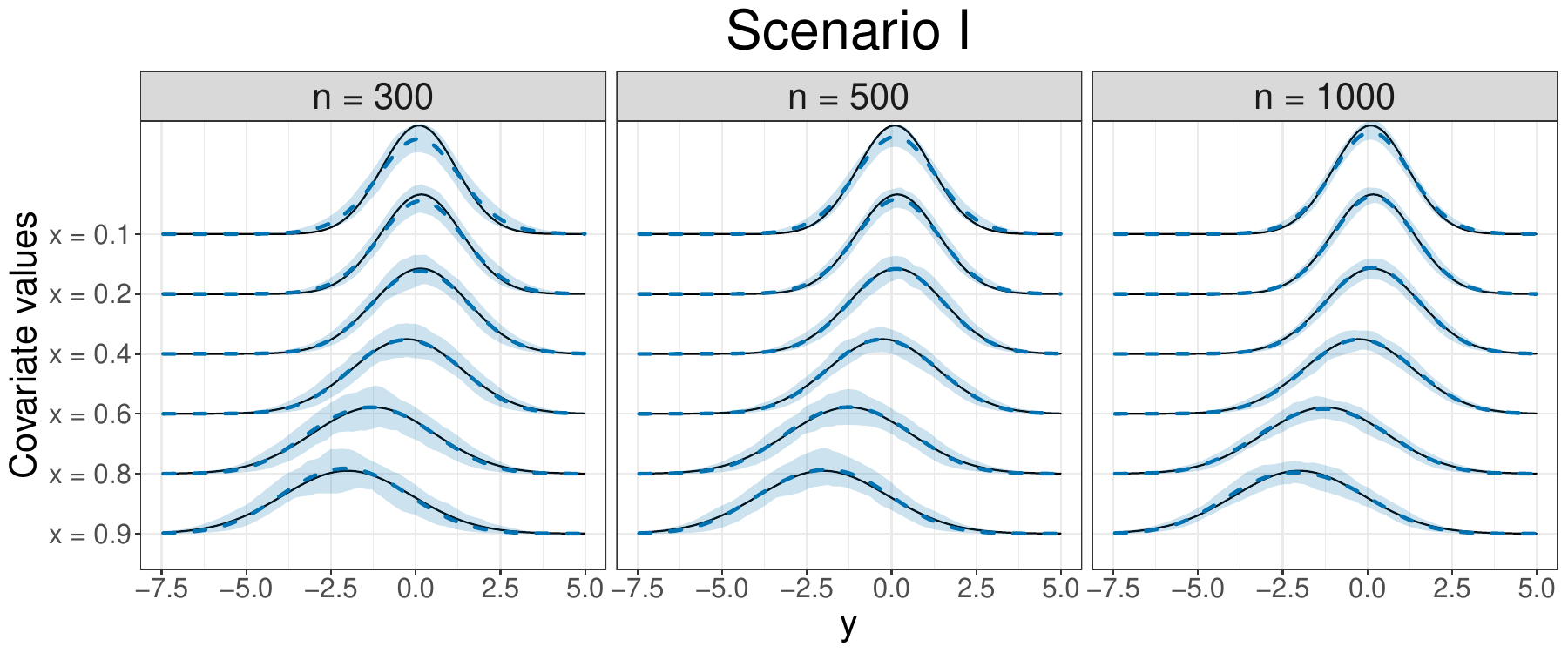}
\includegraphics[scale = 0.37, page = 2]{den_fun_1D.pdf}
\includegraphics[scale = 0.37, page = 3]{den_fun_1D.pdf}
\includegraphics[scale = 0.37, page = 4]{den_fun_1D.pdf}
\end{center}
\caption{For the simulation study and scenarios involving one continuous covariate (Scenarios I, II, III and IV): True (solid black line) and average across the $100$ simulated datasets (blue dashed lines) of the posterior median of the conditional density for $x \in \{0.1, 0.2, 0.4, 0.6, 0.8, 0.9\}$ and different sample sizes ($n$). The shaded areas are bands constructed using the pointwise
2.5\% and 97.5\% quantiles of the 100 posterior medians.}
\label{den_one_cov}
\end{figure}

\begin{figure}[H]
\begin{center}
\includegraphics[scale = 0.5, page = 1]{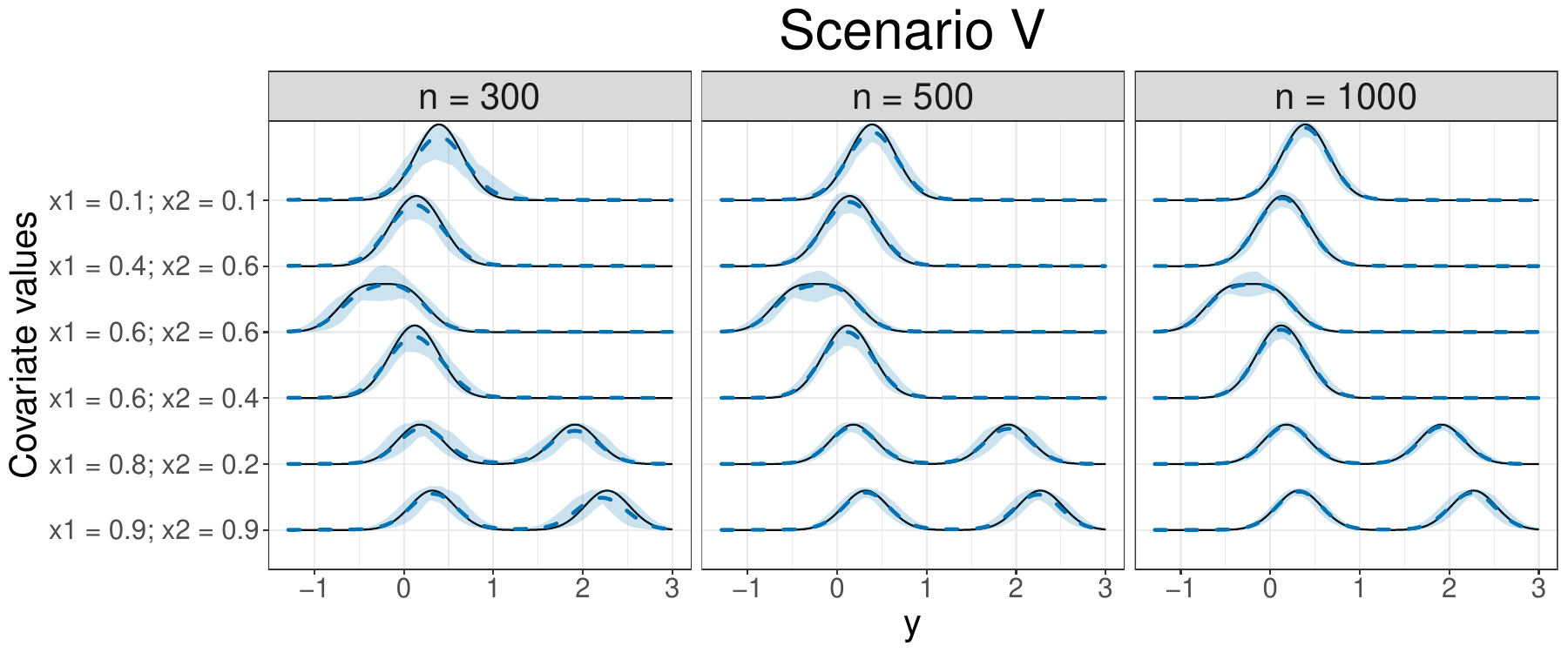}
\includegraphics[scale = 0.5, page = 2]{den_fun_2F.pdf}
\end{center}
\caption{For the simulation study and scenarios involving two continuous covariates and an additive structure (Scenarios V and VI): True (solid black lines) and average across the $100$ simulated datasets (blue dashed lines) of the posterior median of the conditional density for $(x_1, x_2) \in \{(0.1, 0.1), (0.4, 0.6), (0.6, 0.6), (0.6, 0.4), (0.8, 0.2), (0.9, 0.9)\}$ and different sample sizes ($n$). The shaded areas are bands constructed using the pointwise
2.5\% and 97.5\% quantiles of the 100 posterior medians.}
\label{den_two_cov_add}
\end{figure}

\begin{figure}[H]
\begin{center}
\includegraphics[scale = 0.5, page = 1]{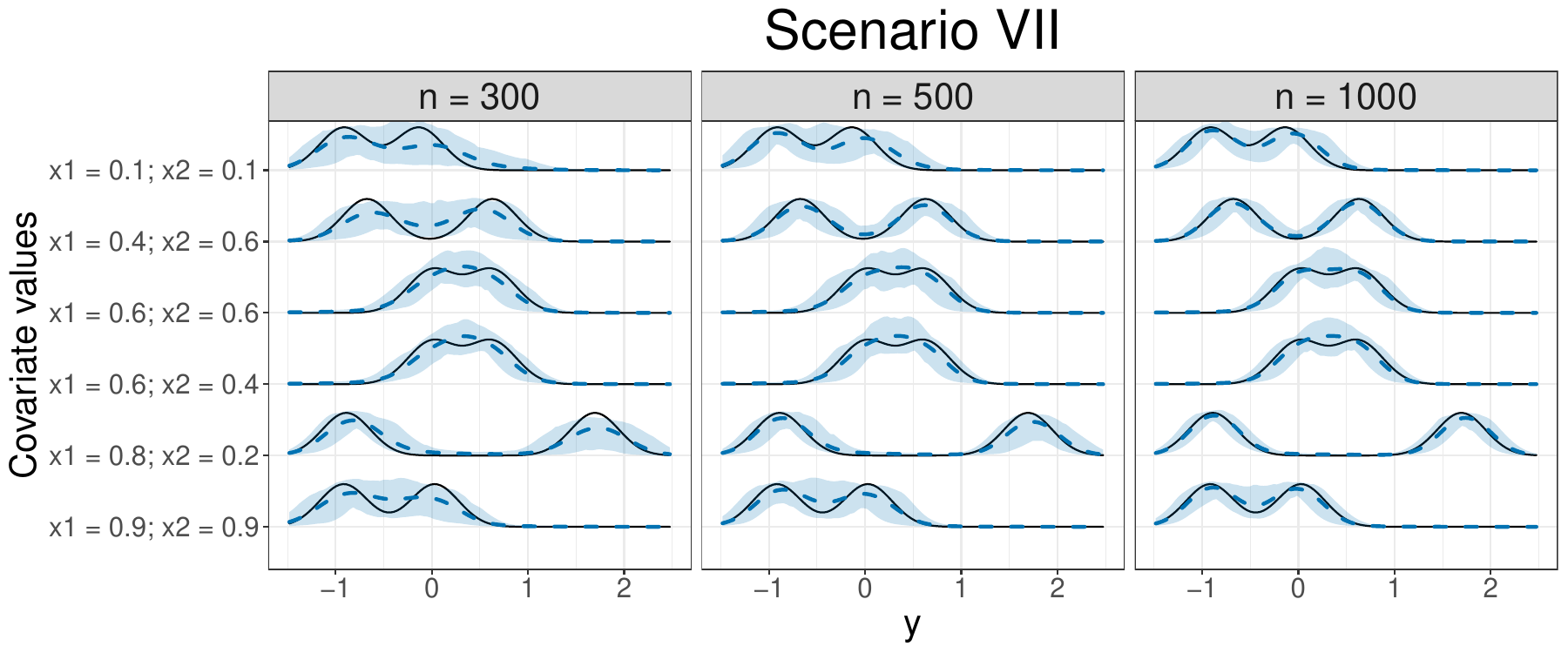}
\includegraphics[scale = 0.5, page = 2]{den_fun_2D.pdf}
\end{center}
\caption{For the simulation study and scenarios involving two continuous covariates and an interaction structure (Scenarios VII and VIII): True (solid black lines) and average across the $100$ simulated datasets (blue dashed lines) of the posterior median of the conditional density for $(x_1, x_2) \in \{(0.1, 0.1), (0.4, 0.6), (0.6, 0.6), (0.6, 0.4), (0.8, 0.2), (0.9, 0.9)\}$ and different sample sizes ($n$).  The shaded areas are bands constructed using the pointwise
2.5\% and 97.5\% quantiles of the 100 posterior medians.}
\label{den_two_cov_int}
\end{figure}

For brevity, the majority of the graphical results are provided in Web Appendix E, and we focus here on the main findings. For Scenarios I to IV, Figure \ref{den_one_cov} shows the average, over the $100$ simulated datasets, of the posterior medians of the predictive conditional density, for $x \in \{0.1, 0.2, 0.4, 0.6, 0.8, 0.9\}$, along with the pointwise $2.5\%$ and $97.5\%$ quantiles of the ensemble of these posterior medians. 
Results for Scenarios V and VI are depicted in Figure \ref{den_two_cov_add} and those for Scenarios VII and VIII are presented in Figure \ref{den_two_cov_int}. For all these four scenarios, which involve two continuous covariates, we consider $(x_1, x_2) \in \{(0.1, 0.1), (0.4, 0.6), (0.6, 0.6), (0.6, 0.4), (0.8, 0.2), (0.9, 0.9)\}$.

We start by discussing the results for Scenarios I to III, all of which involve a single continuous covariate. This is also the case for Scenario IV, whose results are discussed afterwards.  We first note that, as depicted in Figure \ref{den_one_cov},  the conditional densities vary in shape depending on the covariate value; examples include one or two modes and different degrees of asymmetry. This is also true for the other scenarios, as illustrated in Figures \ref{den_two_cov_add} and \ref{den_two_cov_int}. Figure \ref{den_one_cov} shows that our model effectively reconstructs the true profiles of the conditional densities, especially in Scenarios I and II. For Scenario III, which involves covariate-dependent weights, the reconstruction of the profiles of the conditional density is slightly less accurate than in Scenarios I and II, yet the results remain very satisfactory. As expected, across all three scenarios and for all covariates values, the estimates of the conditional density profiles become more concentrated around the true conditional density profiles as the sample size increases. Regarding estimation of the conditional mean, conditional variance, and the three conditional quartiles, the results are presented in Web Figures 2, 7, and 11, respectively. From these figures, it can be observed that our proposed model successfully recovers the true shape of these functionals. However, in the estimates of the conditional variance, some discrepancies appear at the boundaries, particularly for small sample sizes. We recall that in our model, the variance of each component is independent of the covariates. Nonetheless, the variance of the mixture model as a whole does depend on covariates, albeit indirectly, through their effect on the mean of each component (see Web Appendix D). Thus, accurately estimating the conditional variance poses a challenge to our approach. We now shift our focus to Scenario IV, which involves a nonstationary mean function and a variance function which is almost zero for all covariate values between $-2$ and $5$, and beyond this range, it has a quadratic form that increases quite rapidly.  As can be noticed from Figure \ref{den_one_cov}, some profiles of the conditional density, particularly the most peaked ones, are not well recovered. The conditional mean function, shown in Web Figure 2, is quite reasonably recovered. However, the true functional shape of the variance function and of the conditional quartiles are not accurately reconstructed, as shown in Web Figures 7 and 11. In this scenario, flexibility in the mean function is not sufficient to recover the covariate-dependent variance and this also impacts the estimation of the other functionals.  It is important to highlight that model checking procedures can be employed to evaluate the performance of our DDPstar model. This is demonstrated in Web Figures 18 and 19, which respectively display the posterior predictive checks and quantile residuals \citep{Dunn1996} for a single generated dataset under Scenario IV. As evident from these figures, DDPstar is not a suitable approach for this particular scenario.

We now turn our attention to scenarios involving two continuous covariates and an additive structure, Scenarios V and VI. The results depicted in Figure \ref{den_two_cov_add} show that our model successfully recovers the true conditional densities. Results are slightly better for Scenario V, which involves a mixture of two normal regressions and more closely aligns with our model specification. In contrast, Scenario VI comprises a mixture of a skew normal and a $t$ distribution. Results for the other functionals are presented in Web Figure 3 (conditional mean), Web Figure 8 (conditional variance), and Web Figure 12 (conditional quartiles). These results are presented in the form of surfaces, which represent averages of posterior medians across the $100$ simulated datasets. However, we also provide slices of the surfaces for specific covariate values to evaluate the uncertainty associated with our model, as these averages are supplemented with the $2.5\%$ and $97.5\%$ quantiles over the ensemble of posterior medians (see Web Figures 4, 9, 13 and 14). As it can be concluded from all these figures, our method consistently performs satisfactorily in reconstructing the true functionals for these scenarios, with the exception of the variance function, for which it continues to exhibit subpar performance for certain covariate values and small sample sizes. 

We conclude this section by examining the results for the scenarios involving two continuous covariates and an interaction structure, Scenarios VII and VIII. Results for the conditional densities are shown in Figure \ref{den_two_cov_int}. Similarly to Scenarios V and VI, we observe that Scenario VII, which involves a mixture of two normal regressions, yields better results compared to Scenario VIII. However, in contrast to what is observed in Scenarios V and VI, the influence of sample size is more pronounced in this case and larger sample sizes are required to obtain satisfactory results. This is to be expected, as modelling interactions between smooth functions of continuous covariates typically demands substantial sample sizes. Similar conclusions apply to the results for the conditional mean, conditional variance and conditional quartiles, depicted in Web Figures 3 and 5, Web Figures 8 and 10, and Web Figures 15, 16 and 17, respectively.

\section{\large{\textsf{Real data illustrations}}}\label{app_main}
This section demonstrates the broad applicability of the  DDPstar model by applying it to three studies in different fields: toxicology, disease diagnosis and agriculture. For these three applications, comparative assessments are provided against alternative methods that have been applied to the same datasets. In all cases, posterior inference relies on 20000 iterations, after discarding the first 5000 as a burn-in. We employ, for the standardised data, the same hyperparameter values used in the Simulation Study (see Section \ref{scenarios}). We provide the computational times obtained using an Apple Mac mini equipped with an M2 Max chip, 32GB of unified memory, a 12-core CPU and a 30 core GPU.

\subsection{\textsf{Toxicology study}}\label{app_main_tox}
This first study is rooted in epidemiology and has become a benchmark for validating Bayesian nonparametric density regression models \cite[see, e.g.,][]{Dunson2008,Canale2018,Rigon2021}. Its primary objective is to explore the relationship between the concentration of Dichlorodiphenyldichloroethylene (DDE) in maternal serum during the third trimester of pregnancy and the risk of premature delivery \citep{Longnecker2001}. DDE is a metabolite of the pesticide DDT. Despite its potential adverse health effects, DDT is still used to combat malaria-transmitting mosquitoes in areas where malaria is endemic. The dataset consists of 2312 women and the response variable is the gestational age at delivery (GAD), in weeks, and deliveries prior to 37 weeks of gestation are considered as preterm. Interest lies in investigating how the GAD distribution changes with increasing DDE levels, with a particular focus placed on the left tail of the distribution in order to assess the DDE effect on preterm deliveries. 

We apply the DDPstar model to the data $(y_i,x_i) = (\texttt{GAD}_i, \texttt{DDE}_i)$, for $i=1,\ldots,1312$, and consider the following model for the mean of each mixture component (for simplicity, the dependence on the component is omitted)
\[
\mu(\boldsymbol{x}_i) = \beta_0 + f^{\texttt{smooth}}\left(\texttt{DDE}_i\right),
\]
where the smooth effect of DDE is approximated using a cubic B-spline basis of dimension $J = 23$ (we exclude the intercept). For the 20000 iterations, the Gibbs sampler took around 123 seconds. Traceplots and Geweke statistics for the conditional density do not show any lack of convergence of the MCMC chains and effective sample sizes are satisfactory (results not shown). Posterior predictive checks and quantile residuals are provided in Web Figures 21 and 22, respectively, and confirm an accurate fit to the data.

\begin{figure}[H]
\begin{center}
\includegraphics[scale = 0.35]{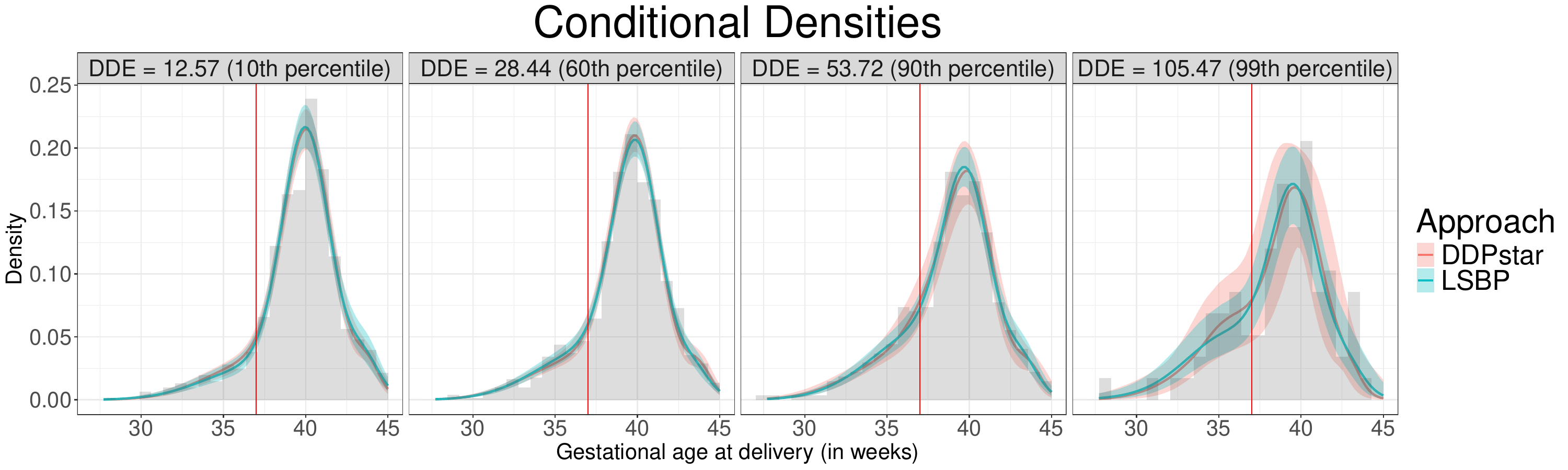}
\end{center}
\caption{Toxicology study: posterior median (solid lines) and 95\% pointwise credible band (shaded
areas) of the conditional density for GAD for selected percentiles of DDE. LSBP stands for the proposal  of \cite{Rigon2021}.}
\label{DDE_densities}
\end{figure}

Web Figure 20 shows the scatterplot of the data along with the regression function (posterior median) and its pointwise $95\%$ credible band. As can be observed, there is a slight decrease in the mean GAD as the DDE levels increase, although for DDE levels above 100mg/L posterior uncertainty is quite substantial due to data scarcity.  In the same figure we also show the variance function and the three conditional quartiles. Figure \ref{DDE_densities} shows ``conditional histograms'', for a range of DDE levels, with the estimated conditional densities superimposed. Similarly to \cite{Dunson2008} and \cite{Rigon2021}, we consider the $0.1$, $0.6$, $0.9$, and $0.99$ quantiles of the DDE. Following \cite{Rigon2021}, the conditional histograms are obtained by grouping the GAD values according to a binning of the DDE with cutoffs at the central values of subsequent quantiles. The estimated conditional densities closely follow the histogram, further indicating a good fit to the data. This figure suggests an increasing left tail of the GAD distribution, which is associated with preterm deliveries, as DDE levels increase. In addition, in Figure \ref{DDE_densities} we also display the posterior inferences obtained using the approach of \cite{Rigon2021}, which allows for covariate-dependent weights. As is apparent, the posterior medians of the two methods are almost indistinguishable. For the $0.9$ and $0.99$ quantiles of the DDE, posterior uncertainty is notably higher with the DDPstar approach, especially for the $0.99$ quantile. This is attributed to: (i) sparsity of the data in this region of the covariate space,  and (ii) our model using a significantly larger number of parameters compared to that of \cite{Rigon2021}. To explore how to use the conditional density to infer conditional preterm probabilities, and following what previous authors did, we compute the following four covariate-specific probabilities: $\Pr(\texttt{GAD}\leq y^{*}\mid \texttt{DDE})$, for $y^*\in \{33, 35, 37,40\}$. Results for both our model and the one from \cite{Rigon2021} are depicted in Web Figure 23. Once again, apart from some relatively minor differences at extreme DDE levels, the two approaches yield similar results.

\subsection{\textsf{Disease diagnosis study}}\label{app_main_roc}
The receiver operating characteristic (ROC) curve stands as the preeminent tool in evaluating the ability of continuous-outcome diagnostic tests to distinguish between individuals with and without a well-defined condition (disease). It is recognised that in many instances, the test outcome and potentially its discriminatory capacity can be affected by covariates (e.g., age,  gender, variations in testing conditions). In such cases, it is advisable to use the conditional or covariate-specific ROC curve, which is an ROC curve that is tailored to a specific covariate value, providing an assessment of the accuracy of the test within the particular subgroup defined by that value \cite[see][and references therein]{Inacio2021_AR}. The covariate-specific ROC curve is defined as
$
\{\left(t, \text{ROC}(t \mid \boldsymbol{x})\right), t \in [0,1]\},
$
where
\begin{equation}
\text{ROC}(t \mid \boldsymbol{x}) = 1 - F_{D}\{F^{-1}_{\bar{D}}\left(1-t \mid  \boldsymbol{x}\right) \mid  \boldsymbol{x}\},
\label{cROC}
\end{equation}
and $F_{D}\left(y \mid \boldsymbol{x} \right) = \Pr\left(Y \leq y \mid \boldsymbol{x}, D = 1\right)$ and $F_{\bar{D}}\left(y \mid \boldsymbol{x} \right) = \Pr\left(Y \leq y\mid \boldsymbol{x}, D = 0\right) $, with $D$ being the binary variable indicating the presence ($D = 1$) or absence ($D = 0$) of disease. Related to the ROC curve, various summary indices serve as measures of a diagnostic test's accuracy. The most commonly used one is the area under the ROC curve (AUC), which, in the conditional context, is defined as  $\text{AUC}(\boldsymbol{x}) = \int_0^1 \text{ROC}(t \mid \boldsymbol{x})\text{d}t$. An AUC of $1$ indicates a test that can perfectly distinguish between diseased and nondiseased individuals, while an AUC of $0.5$ signifies a test with no discriminatory ability.

Our goal is to assess the utility of the body mass index (BMI) as a diagnostic tool for distinguishing individuals with two or more cardiovascular disease (CVD) risk factors (classified as ``diseased'') from those with none or just one CVD risk factor (classified as ``nondiseased''). Since anthropometric measures, including \texttt{BMI}, vary with \texttt{age} and \texttt{sex}, we focus here on estimating age- and sex-specific ROC curves. The dataset comes from a cross-sectional study conducted by the Galician Endocrinology and Nutrition Foundation, consisting of $2840$ individuals aged $18$-$85$ years, with $691$ classified as diseased ($273$ women and $418$ men) and $2149$ as nondiseased ($1250$ women and $899$ men). A detailed description can be found in \cite{Tome09}.

From Equation \eqref{cROC}, it is evident that the estimation of covariate-specific ROC curves can be approached through density regression, i.e., through the estimation of the conditional density function of the diagnostic test (in our case \texttt{BMI}) separately in the diseased and nondiseased population. Here we use the DDPstar model for that purpose. In particular, the following factor-by-smooth interaction models (see also  Web Appendix B) are assumed for the mean of each mixture component in, respectively, the diseased and nondiseased populations (again here the dependence on the component is omitted)
\begin{equation*}
\mu_{D}(\boldsymbol{x}_i)  = \beta_{0D} + \beta_{1D}\mathds{1}_{\{\texttt{sex}_i = \texttt{Women}\}} + f^{D}_{\texttt{sex}_i}(\texttt{age}_i),\quad 
\mu_{\bar{D}}(\boldsymbol{x}_j)  = \beta_{0\bar{D}} + \beta_{1\bar{D}}\mathds{1}_{\{\texttt{sex}_j = \texttt{Women}\}} + f^{\bar{D}}_{\texttt{sex}_j}(\texttt{age}_j),
\end{equation*}
where $\texttt{sex}_i$ (the same applies to $\texttt{sex}_j$) denotes the sex of the $i$th diseased individual (either \texttt{Men} or \texttt{Women}), and $\mathds{1}_{\{\cdot\}}$ is the indicator function, for $i = 1, \ldots, 691$ and $ j = 1, \ldots, 2149$. Note that in our mean formulation, we allow the smooth function of \texttt{age} to be different in men and women and we approximate them using cubic B-spline bases of dimension $J = 23$ (we exclude the intercept). The Gibbs sampler took around 85 and 166 seconds in the diseased and nondiseased populations, respectively. The posterior predictive checks and quantile residuals depicted in, respectively, Web Figures 25 and 26, show a good fit of DDPstar to the data in the two populations.

\begin{figure}[H]
\begin{center}
\includegraphics[scale = 0.55]{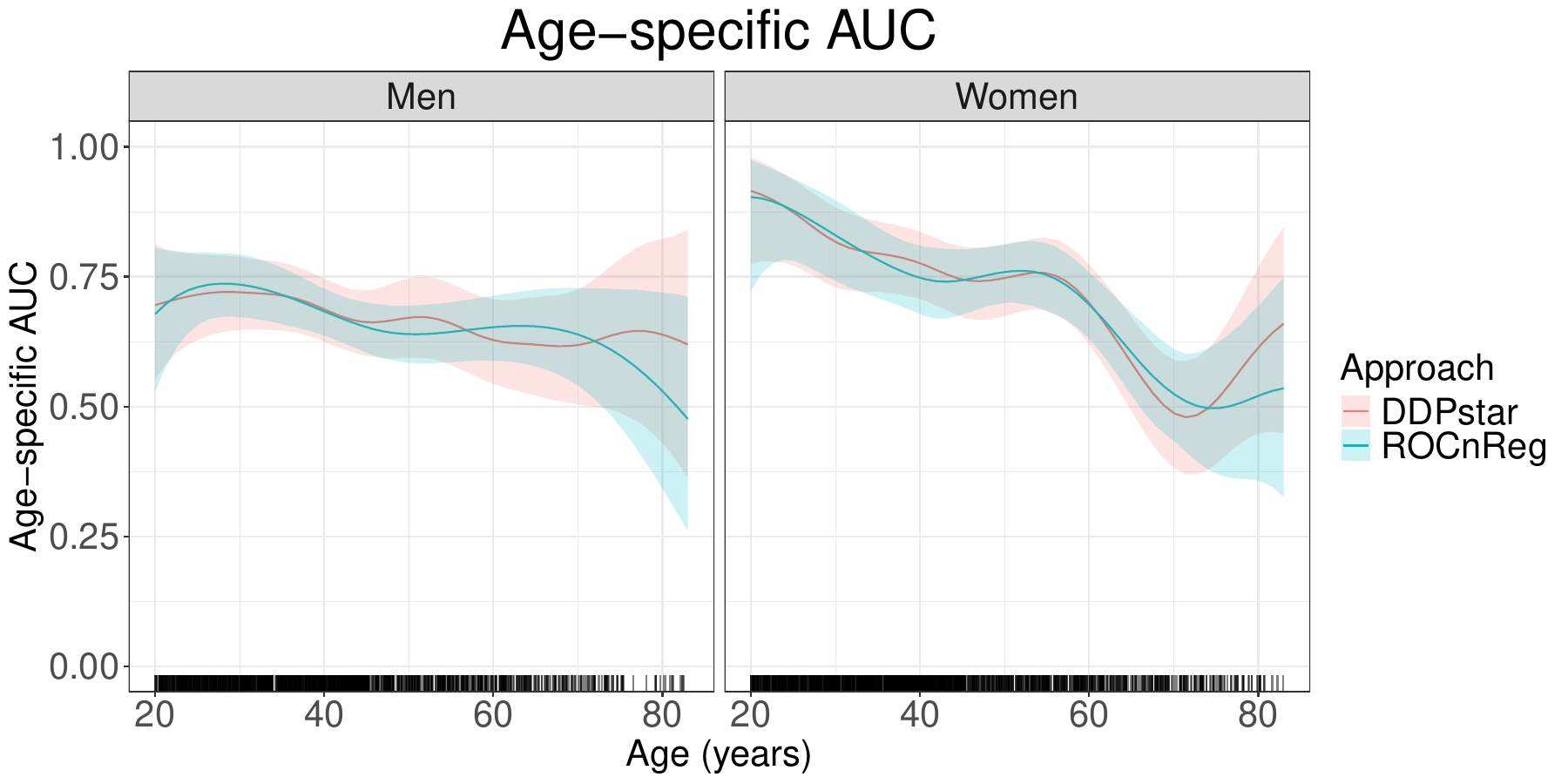}
\end{center}
\caption{Disease diagnosis study: posterior median (solid lines) and 95\% pointwise credible band (shaded
areas) of the age-specific AUC, separately for men (left plot) and women (right plot). DDPstar stands for the proposal presented in this paper and ROCnReg for the one discussed in \cite{Inacio2017} and implemented in \cite{rod_alv_21}.}
\label{ROC_aucs}
\end{figure}

Web Figure 24 depicts the estimated conditional density functions of the BMI in diseased and nondiseased populations. As can be seen, the BMI differs according to age and gender, implying that the accuracy of BMI measurements is also likely to vary across different age and gender groups. This is supported by the results presented in Figure~\ref{ROC_aucs}, which displays the age and gender-specific AUCs. As can be seen, the accuracy of the BMI remains relatively stable across age for men. In contrast, for women, age has a significant influence. Accuracy declines until around $70$ years old, followed by an increase until the age of $85$. 

For comparison, we also obtain age- and sex-specific ROC curves using the proposal described in \cite{Inacio2013}. The same models as for DDPstar are assumed for the mean of each component in the diseased and nondiseased populations, and the smooth functions of age are also approximated using cubic B-spline bases. However, in this approach, results depend heavily on the number of B-spline basis functions (knots are located at the quantiles of the age covariate) since no penalisation is considered. As such, we evaluate, separately for diseased and nondiseased individuals and for men and women, different numbers of B-spline basis functions ($J \in \{3,4,5,6,7\}$), and select the ones that provide the lowest widely applicable information criterion. Note that this implies estimating a total of $20$ ( = $2 \times 2 \times 5$) models. For that purpose, the \texttt{R} package \texttt{ROCnReg} \citep{rod_alv_21} is used and we consider prior distributions that align with those from the DDPstar models. As can be seen from Figure~\ref{ROC_aucs}, which also shows the age and gender-specific AUCs obtained through this unpenalised approach, both DDPstar and ROCnReg approaches yield similar results except for individuals aged $70$ or older. However, uncertainty becomes substantial beyond this age threshold due to limited available data. 

In this application, the primary benefit of DDPstar is that it eliminates the need to examine or assess various B-spline basis dimensions. This leads to a substantial reduction in computational workload, as we only need to fit one model instead of the $20$ considered in the case of ROCnReg.

\subsection{\textsf{Agricultural study}}\label{app_main_agri}
Field trial experiments typically aim to assess the influence of specific genotypes on a targeted trait, which can include crop yield, resistance to pests, or quality measures. These trials often serve commercial purposes, identifying superior genotypes for future commercialisation. However, it is well-known that environmental conditions strongly influence trait/phenotype expression. For instance, soil fertility can greatly affect crop yield, independent of genotype performance. Therefore, in analysing field trials, it is crucial to effectively separate between environmental and genetic influences on phenotypic expression.

\begin{figure}[H]
\begin{center}
\includegraphics[scale = 0.27]{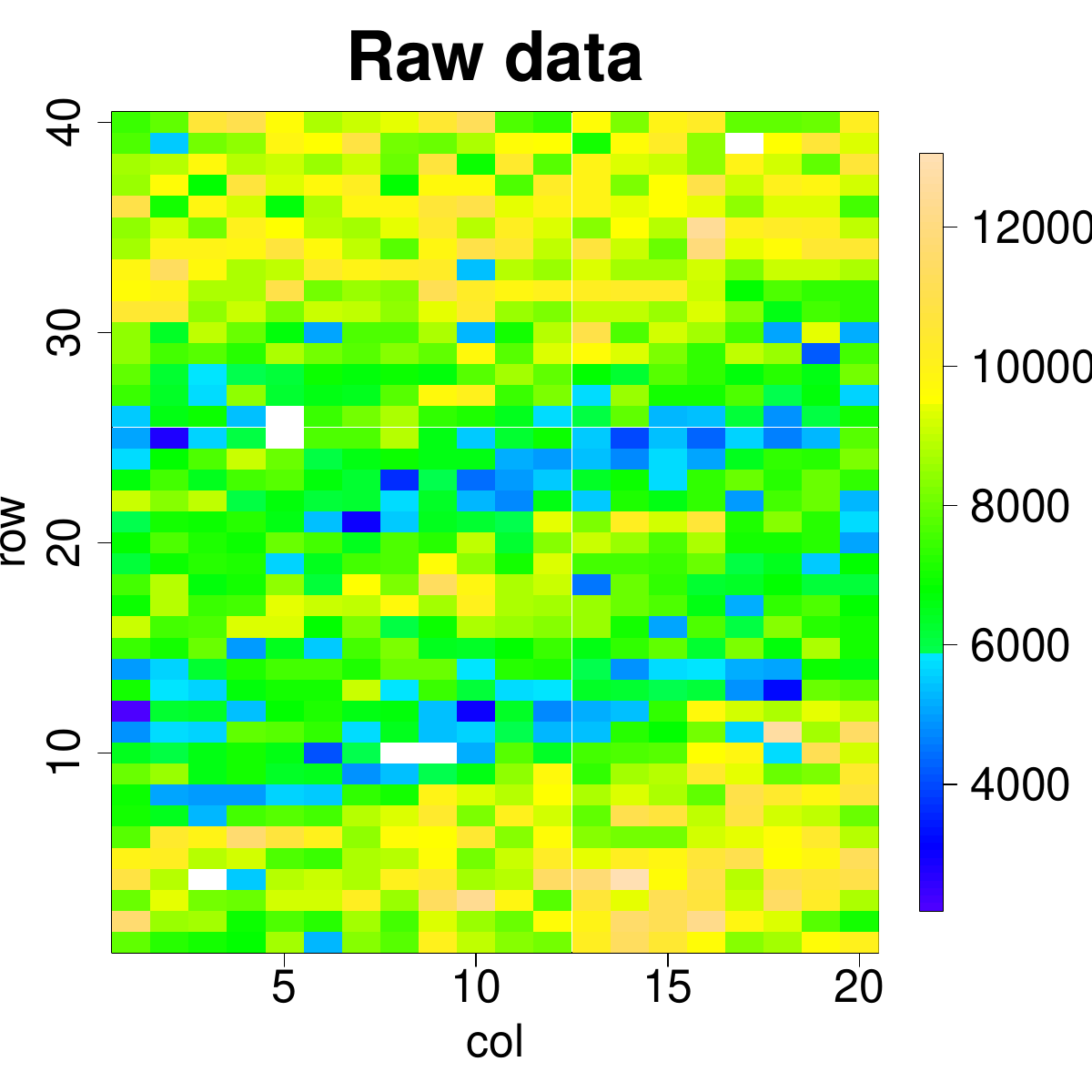}
\includegraphics[scale = 0.27]{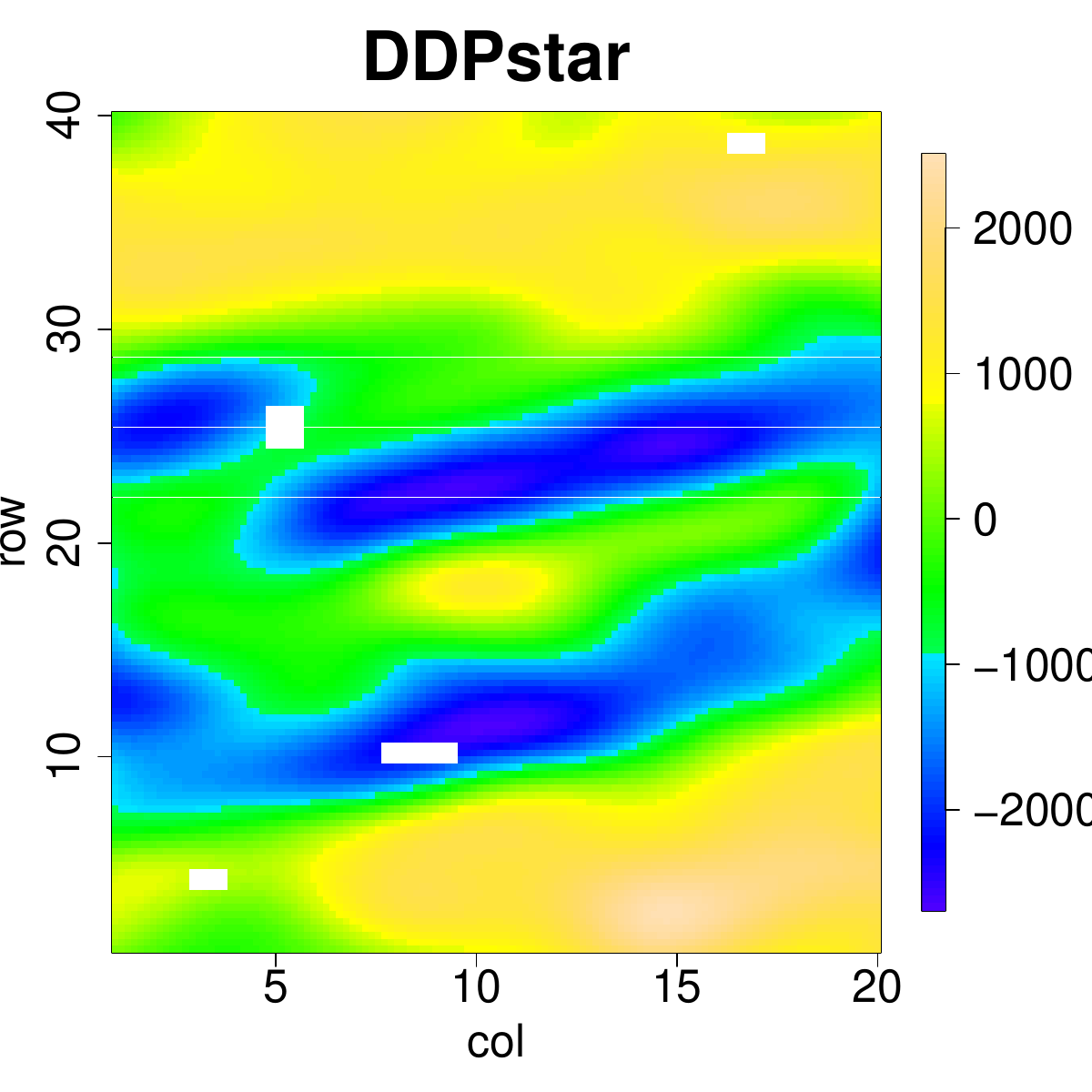}
\includegraphics[scale = 0.27]{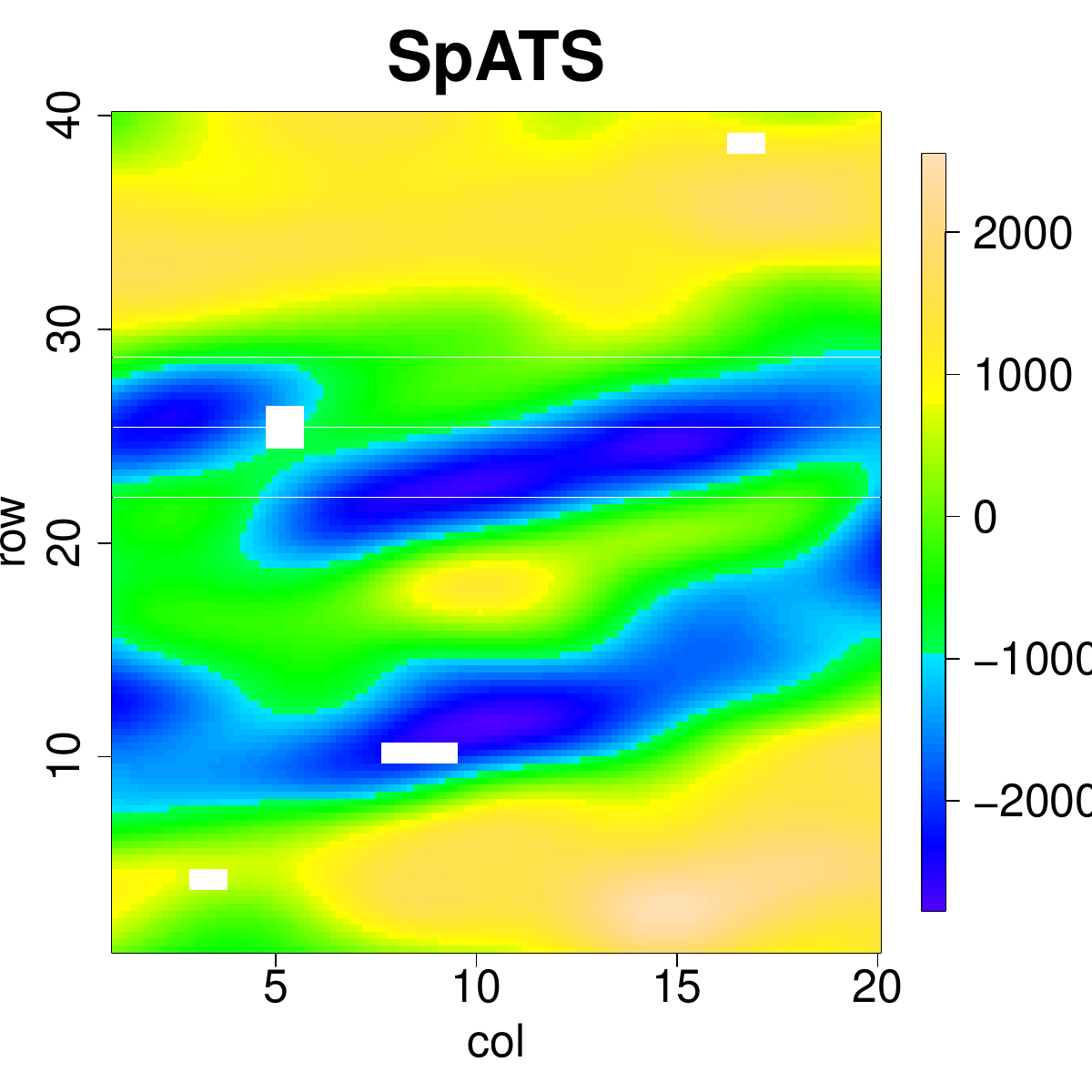}
\end{center}
\caption{Agricultural study: raw data (left), posterior median of the bivariate spatial trend using DDPstar (middle), and estimated spatial trend by means of SpATS (right).}
\label{field_fitted_surfaces}
\end{figure}

As an illustrative example, we examine a Chilean wheat trial described in \cite{Lado2013}. In this study, $384$ diverse wheat genotypes were planted in an $800$-plot field arranged in a $40$-row by $20$-column grid. Genotypes were randomly assigned to plots using an alpha-lattice design with $20$ incomplete blocks, each block containing $20$ genotypes. Two replicates were planted for each genotype. The primary focus here is on grain yield as the trait of interest. The left-hand panel of Figure \ref{field_fitted_surfaces} depicts the observed grain yield in each plot, revealing a pronounced spatial pattern. Because of the randomisation process and the inclusion of replicates, this pattern cannot be attributed to genetics, and it likely arises from environmental factors. Unfortunately, in many cases, there are no covariates available to relate to this effect, leading modelling strategies to depend on the position or spatial coordinates of the plots in the field only.

Recently, \cite{rod_alv_17} proposed a new spatial model, called SpATS, for the analysis of field trials based on using tensor product P-splines to model spatial trends. Here, we adapt the original proposal to our DDPstar model, and specify the mean of each mixture component as a function of available covariates. These  are the genotypes (\texttt{gen}), where the interest is focused, the spatial coordinates of the plots in the field (\texttt{$x_1$} and \texttt{$x_2$} for rows and columns, respectively) as well as the row (\texttt{row}) and column (\texttt{col}) position  of the plots. The model for the mean of each mixture component follows (the dependence on the component's index is omitted)
\[
\mu(\boldsymbol{x}_i) = \beta_0 + \zeta^\texttt{gen}_{g(i)} + \zeta^\texttt{row}_{r(i)} +  \zeta^\texttt{col}_{c(i)} + f(x_{1i}, x_{2_i}),\;\; i = 1,\ldots, 800,
\]
where $f(x_{1}, x_{2})$ is a smooth bivariate function, jointly defined over $x_1$ and $x_2$, that models the spatial effect/trend. We use indices $g$, $r$, and $c$ to refer to genotype, row and column positions, respectively ($g = 1, \ldots, 384$; $r = 1, \ldots, 40$; $c = 1, \ldots, 20$). Accordingly, $\zeta^\texttt{gen}_g$ represents the genotypic effect for the $g$th genotype (similarly for \texttt{row} and \texttt{col}; these effects are typically incorporated to account for the way machines work before and during sowing or planting). To denote the genotype, row and column positions of the $i$th plot, we use $g(i)$, $r(i)$, and $c(i)$, respectively. 

Following \cite{rod_alv_17}, the spatial effect $f(x_{1}, x_{2})$ is modelled by the tensor product of two marginal cubic B-splines bases (excluding the intercept). Since there are twice as many rows ($x_1$ values) as columns ($x_2$ values), we use bases of varying dimensions, with $J_1 = 23$ for $x_1$ and $J_2 = 13$ for $x_2$. Furthermore, we consider $\zeta^\texttt{gen}_{g} \overset{\text{iid}}\sim\,\text{N}(0, \sigma_{\texttt{gen}}^2)$, $\zeta^\texttt{row}_{r} \overset{\text{iid}}\sim\,\text{N}(0, \sigma_{\texttt{row}}^2)$, $\zeta^\texttt{col}_{c} \overset{\text{iid}}\sim\,\text{N}(0, \sigma_{\texttt{col}}^2)$ as prior distributions, i.e., these correspond to random effects, with $\sigma_{\texttt{gen}}^{2}$, $\sigma_{\texttt{row}}^{2}$ and $\sigma_{\texttt{col}}^{2}$ $\sim\,\text{IG}\left(1, 0.05\right)$ (see Web Appendix B). In contrast to the previous studies, here we opt for $L = 5$ components to reduce computational complexity. The Gibbs sampler takes approximately 50 minutes, substantially longer than in the other applications. Posterior predictive checks and quantile residuals are depicted in Web Figures 27 and 28, respectively, and show an extremely accurate fit of DDPstar to the data. 

\begin{figure}[H]
\begin{center}
\includegraphics[scale = 0.35]{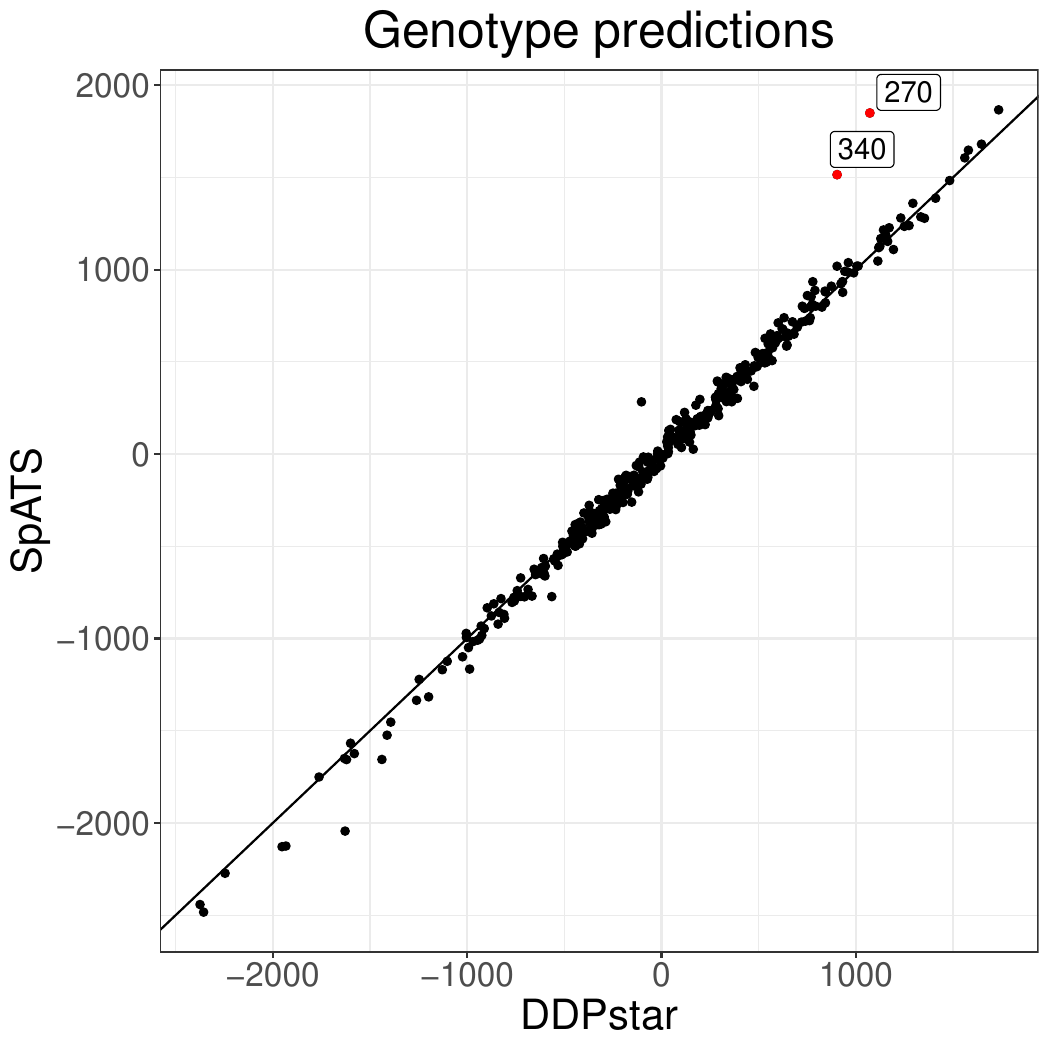}
\end{center}
\caption{Agricultural study: scatter plot of the predicted genotypic effects using SpATS ($y$-axis) against the posterior median of the genotypic effects using DDPstar ($x$-axis). In red two genotypes from the top-performing ones for which SpATS predicts larger effects than DDPstar.}
\label{field_genotype_pred}
\end{figure}

The central panel of Figure \ref{field_fitted_surfaces} displays the posterior median of the bivariate spatial surface. As can be seen, the spatial trend effectively captures the intricate spatial pattern observed in the raw data (left-hand panel of Figure \ref{field_fitted_surfaces}). For comparison purposes, we also fit the model using SpATS. In the context of this paper, a SpATS model corresponds to a model with only one component, estimated through an empirical Bayes approach \cite[for details, see][]  {rod_alv_17}. Figure \ref{field_fitted_surfaces} also depicts the estimated spatial trend using SpATS, revealing minimal differences between both approaches. Figure \ref{field_genotype_pred} presents a comparison between the predicted genotypic effects obtained through our approach and SpATS. Overall, there is a high agreement in the genotypic predictions obtained by both methods. When focusing on the top-performing genotypes, two of them stand out where SpATS predicts larger effects than DDPstar, indicated in red in the figure. Specifically, for genotype number $270$, SpATS and  DDPstar predict effects of $1849.5$ and $1010.7$, respectively. However, it is worth noting that the 95\% credible interval using  DDPstar, $(-380.0, 2471.9)$, encompasses the predicted value obtained from SpATS. A similar result applies to the other genotype (not shown).

All in all, in this specific application, it appears that employing a more intricate model, such as the DDPstar approach, does not yield any apparent advantage compared to SpATS. Nevertheless, these results hold significance as they demonstrate the ability of our approach in handling complex analyses effectively. In fact, the results indicate that only one component of the mixture is effectively occupied, receiving the majority of the weighting (see Web Figure 29), and the interval widths of the $95\%$ credible/confidence intervals for the predicted genotypic effects although larger with DDPstar are still quite comparable (see Web Figure 30).

\section{\large{\textsf{Discussion}}}\label{sec:discussion}
This paper introduced a novel density regression model, DDPstar, which harnesses the flexibility of (i) single-weights DDP mixture models, allowing for a broad class of response distributions and enabling the entire density to smoothly change  with covariates; and (ii) structured additive regression models, 
allowing the accommodation of a variety of covariate effects. Through an extensive simulation study we demonstrated that DDPstar performs effectively even in scenarios where it is misspecified (e.g., in a scenario involving mixture models where the weights of each component depend on covariates). We also provided an example, involving a nonstationary mean function and covariate-dependent variance, where our DDPstar model did not perform well. Although allowing the variance of each normal component to depend on covariates has been shown to be valuable \citep{Villani2009}, assuming a similar additive structure for the variance as for the mean would lead to a model with practically twice the number of parameters per normal component. 
We further illustrated the broad applicability of DDPstar using three real datasets, and its performance was on par with the state-of-the-art approach for each application considered. Our model is implemented in the publicly available \texttt{R} package \texttt{DDPstar} thereby providing a user-friendly and broadly applicable novel approach, and whose performance can be easily assessed through model checking procedures.
 
It is reasonable to inquire about the advantages of assuming an additive structure based on P-splines compared to a Gaussian process \citep{Maceachern2000, Xu2016,Xu2019,Xu2022}. First, there is no closed form expression for the posterior distribution of the hyperparameters of the covariance function of the Gaussian process. These hyperparameters, along with the chosen forms for the covariance and mean functions, are crucial in determining the behaviour of the functionals of interest. Second, if we denote the total number of observations in each mixture component by $n_l$, Gaussian processes require the inversion of a $n_l \times n_l$ matrix in each component, which is computational expensive. In contrast, our model requires only the inversion of a block matrix, which does not depend on the sample size per component, but instead on the number of total B-splines basis functions (see Web Appendix C). We also highlight that although our discussion has focused on the Bayesian nonparametric literature, there are closely related models in the machine learning literature, known as `mixture of experts' \citep[see, e.g.,][Chapter 12, and references therein]{Fruhwirth2019}. 

There are a few limitations to mention. First, in our DDPstar model, allocation to a specific mixture component is independent of the covariates. Therefore, if interest relies on clustering, formulations that allow for covariate-dependent weights should be preferred. Second, except for the agricultural example, our MCMC scheme was quite fast. However our approach does not scale well with either the sample size or the number of covariates. While we do not envision the DDPstar model being used for applications involving a moderate to large number of covariates, variational inference \citep[see, e.g.,][for DPMs]{Blei2006} could be implemented for applications with large sample sizes. Third, although DDPstar provides a flexible formulation for density regression when the response variable is continuous and supported on the real line, it is clearly not suitable for count responses. However, we could easily adapt the proposal of \cite{Canale2011}, based on DP mixtures of rounded Gaussian kernels, to our case. This extension would only involve an additional data augmentation step in our Gibbs sampler scheme. An alternative, less attractive, approach would be to use a transformation of the count data response, e.g., the square root transformation.

A further interesting application of the DDPstar model is that of survival analysis. In this case, considering the logarithm of the event times, and handling right-censored times through a data augmentation step in the Gibbs sampler scheme, DDPstar can be interpreted as a mixture of lognormal accelerated failure time models. In contrast to accelerated failure time and Cox proportional hazards models, the survival curves from different covariate levels would be allowed to cross, a feature that is often appealing in practice. Furthermore, because DDPstar can deal with random effects, the potential inclusion of frailty terms in the aforementioned model would also be interesting to explore.

Supplementary Materials, the \texttt{R} package \texttt{DDPstar} implementing the proposed DDPstar model, as well as the data and \texttt{R}-code needed to reproduce the simulations and data applications can be freely downloaded from \url{https://bitbucket.org/mxrodriguez/ddpstar}.

\section*{\large{\textsf{Acknowledgements}}}
This research was partially funded by the Spanish State Research Agency through the Ram\'on y Cajal grant RYC2019-027534-I.

\bibliographystyle{hapalike}
\bibliography{DDPstar_references}
\end{document}